\documentclass{aa}

\newcommand{\orcid}[1]{\protect\href{https://orcid.org/#1}{\protect\includegraphics[width=8pt]{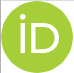}}}
\usepackage{hyperref}
\usepackage{longtable}
\usepackage{float}
\usepackage{placeins}
\usepackage{lineno}
\usepackage{amsmath,bm}
\graphicspath{{./}{figures/}}
\usepackage{graphicx}
\usepackage{placeins}
\usepackage{mathtools, nccmath}
\usepackage{lipsum}
\usepackage{siunitx,etoolbox}
\usepackage{capt-of}
\usepackage{float}
\usepackage[graphicx]{realboxes}
\usepackage[abs]{overpic}
\usepackage{mathrsfs}
\usepackage{txfonts}

\begin{document}

    \title{{\it WI}ggle {\it C}orrector {\it K}it for NIRSp{\it E}c {\it D}ata: WICKED}   
    
    \author{ Antoine Dumont \inst{1} \orcid{0000-0003-0234-3376}
     \and
    Nadine Neumayer \inst{1} \orcid{https://orcid.org/0000-0002-6922-2598}
    \and
    Anil C. Seth \inst{2} \orcid{0000-0003-0248-5470}
    \and
    Torsten Böker \inst{3} \orcid{0000-0002-5666-7782}
    \and
    Michael Eracleous \inst{4} \orcid{0000-0002-3719-940X }
    \and
    Kameron Goold \inst{2} \orcid{0000-0002-7743-9906}
    \and
    Jenny E. Greene \inst{5} \orcid{0000-0002-5612-3427} 
    \and
    Kayhan Gültekin \inst{6} \orcid{0000-0002-1146-0198}
    \and
    Luis C. Ho \inst{7,8} \orcid{0000-0001-6947-5846} 
    \and Jonelle L. Walsh \inst{9} \orcid{0000-0002-1881-5908}
    \and Nora L\"utzgendorf \inst{10} \orcid{0000-0001-6126-5238}
    }
    
    \institute{Max Planck Institute for Astronomy, K{\"o}nigstuhl 17, 69117 Heidelberg, Germany\\ \email{antoine.dumont.neira@gmail.com} \and Department of Physics and Astronomy, University of Utah\\ 115 South 1400 East, Salt Lake City, UT 84112, USA \and European Space Agency, Space Telescope Science Institute, Baltimore, Maryland, USA \and Department of Astronomy \& Astrophysics and Institute for Gravitation and the Cosmos, The Pennsylvania State University, 525 Davey Lab, University Park, PA 16802 \and Department of Astrophysical Sciences, Princeton University, 4 Ivy Lane, Princeton, NJ 08544, USA \and Department of Astronomy, University of Michigan, 1085 S. University Avenue, Ann Arbor, MI 48109, USA \and Kavli Institute for Astronomy and Astrophysics, Peking University, Beijing 100871, China \and Department of Astronomy, School of Physics, Peking University, Beijing 100871, China \and George P. and Cynthia Woods Mitchell Institute for Fundamental Physics and Astronomy and Department of Physics and Astronomy, Texas A\&M University, College Station, TX 77843, USA \and European Space Research and Technology Centre, Keplerlaan 1, 2200 AG Noordwijk, The Netherlands
    }

    \abstract
    {
    {\it Context:} The point-spread function of the integral-field unit (IFU) mode of the Near-Infrared Spectrograph (NIRSpec) detector of JWST is heavily under-sampled. The resampling of the spectra into a 3D data cube creates resampling noise seen as low-frequency sinusoidal-like artifacts, or ``wiggles''. These artifacts in the data are not corrected in the JWST data pipeline, and significantly impact the science that can be achieved at a single-pixel level. 
    
    {\it Aims:} Here we present the tool ``{\it WI}ggle {\it C}orrector {\it K}it for NIRSp{\it E}c {\it D}ata'' ({\sc WICKED}), designed to remove these artifacts. While fully characterizing wiggles requires forward modeling of the instrument response, {\sc WICKED} offers a faster, computationally efficient alternative using an empirical correction.
    
    {\it Methods:} {\sc WICKED} uses the  Fast Fourier Transform to identify wiggle-affected spaxels across the (IFU) data cube. Spectra are modeled with a mix of integrated aperture and annular templates, a power-law, and a second-degree polynomial, avoiding high-degree polynomials that distort spectral features. Our correction works across all medium and high-resolution NIRSpec gratings: F070LP, F100LP, F170LP, and F290LP.
    
    {\it Results:}{ \sc WICKED} can recover the true overall spectral shape up to a factor of 3.5$\times$ better compared to uncorrected spectra. It recovers the equivalent width of  absorption lines within 5\% of the true value \textemdash $\sim$3$\times$ better than uncorrected spectra and $\sim2\times$ better than other methods. {\sc WICKED} significantly improves kinematic measurements, recovering the line-of-sight velocity (LOSV) within 1\% of the true value \textemdash more than $100\times$ better than uncorrected spectra at S/N $\sim$40 . The superior wiggle-removal capabilities of {\sc WICKED} also reduces the LOSV uncertainties by $\sim$50\% compared to other methods. As a case study, we applied {\sc WICKED} to G235H/F170LP IFU data of the elliptical galaxy NGC~5128, finding good agreement with previous studies. In wiggle-affected regions, the uncorrected spectrum showed stellar LOSV and velocity dispersion differences compared to the {\sc WICKED}-cleaned spectrum, of $\sim$17$\times$ and $\sim$36$\times$ larger than the estimated uncertainties, respectively.

    {\it Conclusions:}{Wiggles in NIRSpec IFU data can significantly distort the overall spectral shape, bias line measurements and kinematics to values larger than the expected uncertainties for uncorrected spectra. \sc WICKED} is a robust, user-friendly solution for mitigating wiggles in NIRSpec data. Unlike other methods, it minimizes residual artifacts, enabling precise single-pixel studies, enhancing JWST’s potential for groundbreaking discoveries in galaxy kinematics and early universe studies.  }

    \keywords{Astronomical instrumentation, methods and techniques -- Galaxies: kinematics and dynamics, elliptical and lenticular, nuclei}
    \maketitle
    
\section{Introduction\label{sec:intro}} 

Three years since the launch of the James Webb Space Telescope (JWST) \citep{JWST}, its unprecedented gain in sensitivity in the near- and mid-infrared has already revolutionized the study of the early universe. 
JWST is equipped with two integral field unit (IFU) spectrographs, NIRSpec \citep{NIRSpec_instrument} and MIRI \citep{MIRI_instrument}, which provide spectral information across the field-of-view, enabling studies of kinematics and chemical abundances across the field of view. However, both IFU units of JWST are spatially under-sampled, staying below the desired Nyquist sampling at any wavelength, which creates significant resampling noise modulations \citep{Smith2007,Law2023}.
NIRSpec, in particular, is the IFU mode most affected by under-sampling of the point-spread function (PSF), with a Full Width Half Maximum (FWHM) at 3 microns equal to its pixel size of 0.1” \citep{Ruffio2024}. As a result, prominent low-frequency (5-60 [1$/\mu m$]) PSF artifacts commonly referred to as ``wiggles'', are present in many NIRSpec IFU observations, currently without correction in the JWST pipeline.  With JWST breaking records for proposal submissions and NIRSpec being the most demanded instrument\footnote{Source:  \href{https://www.stsci.edu/contents/news/jwst/2024/jwst-observers-break-their-own-record-for-astronomical-proposal-submissions}{ STScI JWST website.}}, addressing these artifacts has become a priority. 
The ideal approach for correcting these artifacts would involve forward modeling of mock data to accurately simulate the NIRSpec instrument. These simulations could then be compared to observations to identify and subtract the wiggles. Forward modeling has been done for the NIRSpec multi-object spectroscopic (MOS) mode for a sample of high-redshift galaxies in the JWST Advanced Deep Extragalactic Survey (JADES) \citep{DeGraaff2023}, but it is currently unavailable for the IFU mode. Forward modeling is computationally expensive and challenging to implement due to the complexity of the JWST NIRSpec PSF with \citet{DeGraaff2023} and \citet{Ruffio2024} being the only examples available so far. Additionally, generating accurate simulations to account for the unique characteristic of every data set is also a challenge. Consequently, empirical, post-facto correction methods offer a more practical and efficient alternative.
Previous studies using NIRSpec IFU data have often relied on spatially binning the data to apertures equal to or larger than the PSF \citep[e.g.][]{Bianchin2024}, particularly for studies of point sources where these artifacts are more pronounced. Others have addressed the issue by simply ignoring and masking out the affected spaxels \citep[e.g.][]{Donnan2024}. Two groups have attempted to develop fitting routines to remove these wiggles: one using an integrated spectrum template \citep{Perna2023}, and the other a single power-law \citep{Doan2024}, to model the spectrum across the IFU's field of view. The wiggles are then identified as the difference between the data and the model.
The wiggles in the spectrum display a sinusoidal pattern with varying amplitude and frequency across the wavelength range. While the amplitude and phase of the wiggles for each spaxel in the IFU are different, \citet{Perna2023} found that the wiggle frequencies are not random but correlate with wavelength. This correlation appears consistent across all spaxels in the field of view and can be used to model the wiggles as a series of sine functions with a frequency determined by the wavelength.
The method by \cite{Perna2023} demonstrated the feasibility to remove wiggles using an empirical post-facto approach, and is currently the best available tool to remove these artifacts from NIRSpec IFU data. However, it has several limitations: (1) it can significantly alter the spectral shape of the continuum, (2) it lacks a robust method for flagging affected spaxels across the field-of-view, and (3) it is restricted to the G235H/F170LP \& G395H/F290LP configuration. 

To address these challenges, we developed {\sc WICKED} (Wiggle Corrector Kit for NIRSpec Data), a user-friendly {\sc Python} Class designed to remove resampling noise artifacts or wiggles from NIRSpec IFU data for all medium and high-resolution gratings; F070LP, F100LP, F170LP, and F290LP. {\sc WICKED} applies the Fast Fourier Transform to identify and flag affected spaxels while avoiding the high-degree polynomials used by \citet{Perna2023} to model the continuum of the spectrum which affects the integrity of the data. 

The paper is organized as follows: In Section~\ref{sec:data_reduction} we describe the data reduction for the NIRSpec IFU data of two stars, as well as for the elliptical galaxy NGC~5128 (also known as Centaurus~A), used as test cases. In Section~\ref{sec:WICKED} we give a detailed description of the {\sc WICKED} algorithm. Section~\ref{sec:wicked_tests} shows the results of several tests developed to quantify {\sc WICKED}'s performance. Section~\ref{sec:Kinematics_CenA} shows the results of the stellar and gas kinematics for the NIRSpec F170LP IFU data of NGC~5128, corrected for wiggles using {\sc WICKED}. Finally, in Section~\ref{sec:Conclusion} we present a summary of the methodology and the results for {\sc WICKED}, before discussing the findings for the kinematics of NGC~5128.

\section{Data}
In this work, we used JWST NIRSpec IFU archival observations of the two stars 2MASS J17571324+6703409, 2MASS-J15395077\-3404566, and the nucleus of NGC5128. The high resolution IFU mode of NIRSpec can obtain spectra in the wavelength range of $0.6-5.3$ $\mu m$ for a $3.1\arcsec \times 3.2\arcsec$ field of view and a spectral resolution of $R\approx2700$ \citep{NIRSpec_Boker}. We used the stellar spectra of stars for testing the performance of WICKED, and the nucleus of NGC5128 as a practical example of the science that can be achieved after correcting for wiggles in the NIRSpec IFU spectra. The observations for the A-type star 2MASS J17571324+6703409 (hereafter ``J17571324'') are part of Cycle 2 program PID3399 (PI: Marshall Perrin) taken with a 16-dither pattern in the high-resolution modes G140H/F100LP, G235H/F170LP and G395H/F290LP, with a total integration time of 3034 s, 3968 s and 4901 s respectively. Observations for the M-type star 2MASS-J15395077-3404566 (hereafter: ``J15395077'') were taken for Cycle 1 program PID1364 (PI: Misty C. Bentz) using of four-dither pattern with a total integration time of 171.2 s in the high resolution G235H/F170LP configuration. Finally, observations of the nucleus of NGC~5128 were obtained for a GTO program PID1269 (PI: Nora Luetzgendorf) using a four-point dither pattern with a total integration time of 933.7 s per configuration in the high-resolution modes G235H/F170LP and G395H/F290LP, with one leakcal image taken per configuration with a total integration time of 233.5 s. 

\subsection{NIRSpec IFU Data Reduction}
\label{sec:data_reduction}

The NIRSpec IFU data were reduced using the JWST data pipeline version 1.12.5 and context file ``jwst$\_$1256.pmap'' 
\citep{Bushouse2023}.The data reduction was identical for J17571324, J15395077, and NGC~5128, except that for NGC~5128 a leakcal image was used. We started the reduction by running the {\rm Detector1Pipeline} module of the JWST pipeline on all raw {\it uncal.fits} files. We used the patch {\sc snowblind} \citep{snowblind} during {\rm Detector1Pipeline} to correct for large cosmic ray hits. For this,  we saved the ``jump.fits'' files during the {\sc Detector1Pipeline} and passed them to {\sc snowblind} which detected and masked pixels affected by cosmic-ray hits. 
The clean ``jump.fits'' are passed again to {\rm Detector1Pipeline} to run the remaining steps until obtaining a ``rate.fits'' file. The count-rate files were then processed using the {\rm Calwebb\_spec2} module with default parameters, but using the NSClean \citep{NSClean} step (now incorporated into the JWST pipeline) to correct for correlated noise. Finally, the resulting ``cal.fits'' files were resampled and co-added using the ``drizzle'' weighting into a cube using the {\rm Calwebb\_spec3} with a spaxel size of $0\farcs1$ and the instrument alignment. A {\sc Jupyter Notebook} template of our JWST pipeline sequence can be found at \url{https://github.com/antoinedumontneira/NIRSpec-Ppipeline-Template}.

All three resulting cubes present substantial sinusoidal patterns in their spectra at the single spaxel level, due to resampling noise caused by the undersampling of the PSF \citep{Law2023}. Wiggles are particularly noticeable near the cores of compact sources such as stars, active galactic nuclei (AGN), and quasars. The effect is further amplified in data cubes with better spatial sampling (i.e., smaller spaxel sizes like $0.05\arcsec$) and those constructed using the ``drizzle'' weighting instead of ``emsm'' \citep{Perna2023}. To correct these artifacts we developed the {\sc Python} routine {\sc WICKED} which we describe in more detail below. 

\section{The WICKED code}
\label{sec:WICKED}
In this section, we give a detailed description of the ``Wiggle Corrector Kit for NIRSpec Data'' ({\sc WICKED}). The workflow has three main steps that run one after the other. First, {\sc FitWigglesCentralPixel} fits the spectrum of the brightest spaxel to constrain the frequency correlation of the wiggles. Second, {\sc get\_wiggly\_pixels} flags spaxels across the (IFU) data cube  significantly affected by wiggles using the Fast Fourier Transform based on the frequency range of the wiggles found in the first step. Third, {\sc FitWiggles} fits the wiggles in the flagged spaxels, removes them, and saves the cleaned spectra as a new datacube. Steps 2 and 3 in {\sc WICKED} are set to run in parallel on multiple CPUs, allowing each spaxel to be handled separately and making the process faster. {\sc WICKED} is available to the public and can be downloaded for free. We have made a well-documented {\sc Jupyter Notebook} example in the {\sc GitHub} repository \url{https://github.com/antoinedumontneira/WiCKED} to help users get started. 

Subsection~\ref{subsec:wiggle_model} gives an in-depth description of how {\sc WICKED} fits the spectra to create a model of the wiggles. Similarly, Subsection~\ref{subsec:flagging_pixels} describes how the wiggles are identified across the different spaxels in the (IFU) data cube using the Fast Fourier Transform.

\subsection{Modeling the wiggles}
\label{subsec:wiggle_model}

The first step in {\sc WICKED} 
is to create a wiggle-free model of the spectrum. This model is then subtracted from the observed spectrum to identify the wiggles. 
Wiggles are PSF artifacts that depend on the shape of the source; the more point-like the source, the more pronounced the wiggles in the spectrum. Therefore, in {\sc WICKED} we first characterize the wiggles in the brightest spaxel of the data cube. This pixel should have the strongest wiggles, and at the same time, it has the highest signal-to-noise ratio ({\it S/N}). This spaxel serves as the reference for constraining the frequency of the wiggles in the spectrum. {\sc WICKED} has a built-in method {\sc get\_center} to identify the brightest spaxel.

\begin{figure*}
    \centering
    \includegraphics[width=.9\linewidth]{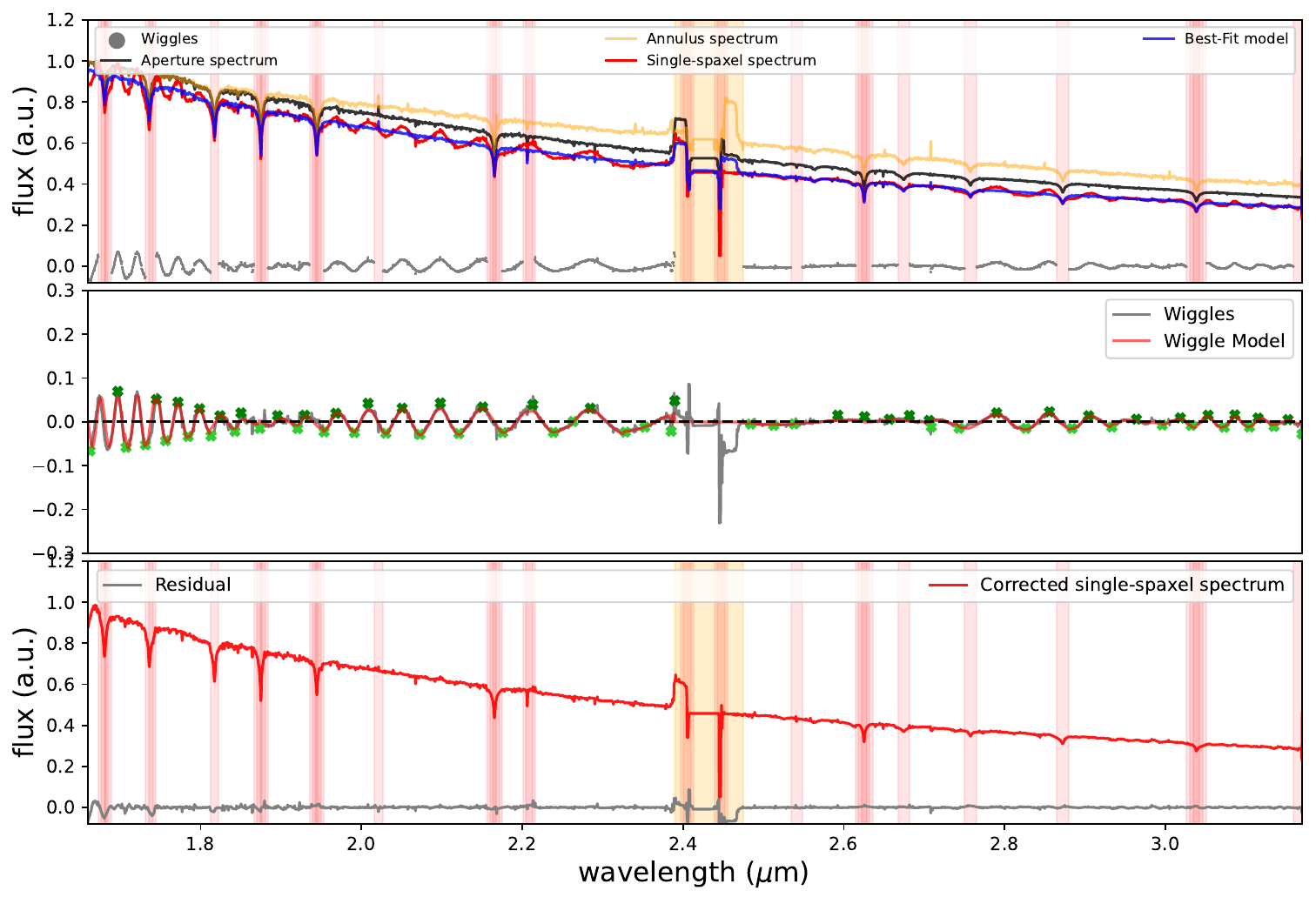}
    \caption{ Correction of the brightest spaxel in the F170LP cube of the A-star J1757132 using the {\sc FitWigglesCentralPixel} step in {\sc WICKED}. The top panel shows the original spectrum (red) with prominent wiggles. The aperture (3-pixel radius) and annular integrated (3 to 5 pixel annular radius) spectra are displayed in gray and yellow, respectively, with the best-fit model shown in blue. Subtracting this smooth model from the single spaxel data produces the wiggle spectrum (middle panel, gray). {\sc WICKED} identifies peaks and valleys in the wiggle spectrum (green crosses) and fits a wiggle model (red), which is subtracted to produce the final wiggle-free spectrum (bottom panel, red). pink vertical lines mark masked regions, while the yellow area highlights the gap between the NRS1 and NRS2 detectors.
    }
    \label{fig:central_pix_example}
\end{figure*}

{\sc WICKED} fits the spectrum using two spectral templates obtained from an aperture ($T_{1}$) and an annular extraction ($T_{2}$), plus a power-law continuum and a second-degree polynomial. The final model is:
\begin{equation}
    w_{1}\cdot T_{1} + w_{2}\cdot T_{2} + w_{3}\cdot(a_{1}\cdot\lambda^{a_{2}}+a_{3}) + b_{1}\cdot\lambda + b_{2}\cdot\lambda^{2}+b_{3}
\end{equation}
where $w_{1,2,3}$ are the weights of the aperture and annular templates ($T_{1}$ and $T_{2}$, respectively) and the power-law, $a_{1,2,3}$ are the power-law parameters, and $b_{1,2,3}$ define the polynomial. The two integrated spectral templates (with user-defined radii) help to preserve spectral features and to model line strength and spectral shape variations across the (IFU) data cube. Since wiggles are spatially correlated, integrating over multiple spaxels effectively removes them from the integrated spectral templates. The additional power-law continuum and second-degree polynomial help to model any difference between the spectral templates and the observed spectrum. This is effective because many physical processes (black hole accretion, dust extinction, background, etc.) can be approximated by a power law, especially over as short wavelength range as that of a NIRSpec passband. We found that adding an additional second-degree polynomial to our models improved the quality of the fits. This polynomial is mostly used to fit ``bump'' like features in the spectrum.  The approach described here avoids using a high-degree polynomial as in \citet{Perna2023}. In our tests, high-degree polynomials yielded unpredictable fits, significantly affecting the continuum shape in some parts of the spectrum. The high-degree polynomial would also ``partially'' fit the wiggles, jeopardizing the performance of the code in some cases. 

Once {\sc WICKED} obtains a good fit of the spectrum, it is subtracted from the data to obtain a wiggle spectrum. The wiggles show a sinusoidal shape with varying amplitudes and frequencies across the wavelength range. Hence, the wiggles cannot be modeled as a single sine function but as a series of different sines of the form $y(\lambda) = A\times cos(2\pi f_{\lambda}\lambda + \frac{\pi}{2}\phi)$.  We subdivide the wiggle spectrum on the basis of its peaks and valleys helping to constrain their frequencies. We found that dividing the wiggle spectrum based on a ``rolling-window'' of a random (fixed) width as in \citet{Perna2023} resulted in sub-optimal wiggle removal; a window of a single width cannot properly account for the diversity of wiggles. For example, if the random width is shorter than the wavelength of the modulation, the code would model a series of smaller sinusoidals of high frequency. In contrast, if the width was larger than the wavelength of the modulation, it would enclose wiggles of different frequencies at the same time, resulting in a worse fit.

After, the wiggle spectrum is split into different slices based on its peaks and valleys. Each slice is fitted with a sinusoidal model $y(\lambda) = A\times cos(2\pi f_{\lambda}\lambda + \frac{\pi}{2}\phi)$. This process is repeated {\it N} times (user-defined, with a default of 15). For each slice, we save the best-fit frequency $f_{\lambda}$ and the central wavelength. A five-degree polynomial is then fit to the $f_{\lambda}$ and central wavelength in each iteration to constrain the wiggle frequency $f_{\lambda}-\lambda$ relation. This is then used as a prior for finding the best $f_{\lambda}$ in the subsequent iterations. The polynomial fit is updated with the new frequencies in each iteration if the new frequencies pass a chi-square threshold set by the first iteration. This ensures that only good fits are saved, preventing poor fits from negatively impacting the polynomial fit of the $f_{\lambda}-\lambda$ relation. Finally, in the last iteration, the best-fit wiggle model is subtracted from the brightest spaxel, correcting for the undesired wiggles.

Figure~\ref{fig:central_pix_example} shows the output of the {\sc FitWigglesCentralPixel} step in {\sc WICKED} that fits the data and constrains the $f_{\lambda}-\lambda$ relation of the wiggles for the brightest spaxel of the data, in this case of the A-star J1757132. The top panel of Figure~\ref{fig:central_pix_example} shows in red the spectrum of the brightest spaxel for J1757132, with the two integrated spectrum templates in gray and yellow. The best-fit model built from the spectral templates, the power-law and the polynomial is shown in blue. The middle panel of Figure~\ref{fig:central_pix_example} shows the wiggle spectrum (gray) and the best wiggle model (red) obtained by {\sc WICKED}. Highlighted as green crosses are the identified peaks and valleys used to split the wiggle spectrum. The bottom panel shows the corrected spectrum (red) and the residuals (gray) between the best-fit model and the corrected spectrum. The vertical lines indicate masked regions, which are excluded during the fitting of wiggles, and the yellow region the wavelength gap between the two NIRSpec detectors. 

The process described here is repeated for each spaxel of the data cube (inside a search radius, see more details in Subsection~\ref{subsec:flagging_pixels}), with an adjusted second-degree polynomial and power-law coefficient for each. Additionally, the $f_{\lambda}-\lambda$ relation found for the brightest spaxel is used as a prior to fit for $f_{\lambda}$. Contrary to the method of \citet{Perna2023}, where the $f_{\lambda}-\lambda$ relation is the same for all spaxels across the data cube, we allow the $f_{\lambda}-\lambda$ to change for each spaxel. 

\subsection{Flagging spaxels with wiggles}
\label{subsec:flagging_pixels}

The identification of spaxels in the data cube affected by wiggles is performed in the {\sc FindWiggles.get\_wiggly\_pixels} step in {\sc WICKED}. In this step {\sc WICKED} calculates the Fast Fourier Transform of the wiggle spectrum using the {\sc SciPy} package {\sc scipy.fft}. The resulting Fourier spectrum ($\mathscr{F}_{Spec}$) is divided into two sections; one part dominated by wiggles and the rest of the spectrum. The part of the Fourier spectrum dominated by wiggles is defined based on the frequencies found during the {\sc FitWigglesCentralPixel} step for the brightest spaxel. 
The frequencies of the wiggles depend on the compactness of the source, typically ranging between $5-60 \: [\mu m^{-1}]$. The mean amplitude and standard deviation for these two parts of the Fourier spectrum are then compared to identify spaxels with wiggles. The mean amplitude plus the standard deviation (a 1-$\sigma$ value) for the part of the spectrum not dominated by wiggles give us information about the amplitude of the stellar features and the noise in the spectrum. We use this as a benchmark and compare it to the mean amplitude of the Fourier spectrum at frequencies dominated by wiggles. For spaxels affected by wiggles, the ratio of these two Fourier amplitude values (hereafter ``Fourier ratio'') is larger than one. The Fourier ratios is:

\begin{equation}
    F_{ratio} = \frac{Mean(\mathscr{F}_{Spec} (\leq f_{wiggles})) }{ Mean(\mathscr{F}_{Spec}(>f_{wiggles})) + \sigma(\mathscr{F}_{Spec}(>f_{wiggles}))} 
\end{equation}

\begin{figure*}[h!]
    \centering
    \includegraphics[width=0.8\linewidth]{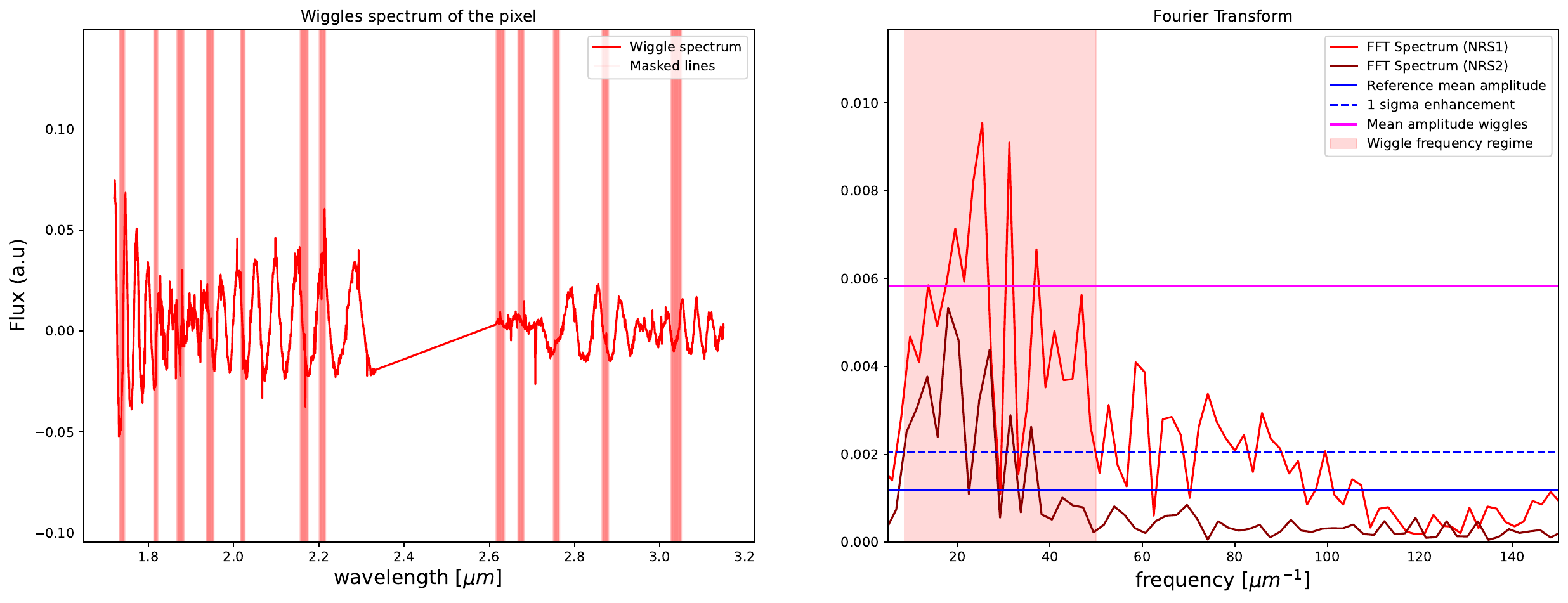}
    \caption{The Fast Fourier Transform is used in {\sc WICKED} to flag spaxels in the data cube affected by wiggles. The built-in method {\sc plot\_wiggle\_FFT} in WICKED allows for manual examination of the spectrum and its Fourier transform for a specific spaxel in the datacube. 
    Left panel shows the wiggle spectrum, created by subtracting the spectrum from the best-fit model. Right panel shows the Fourier transform of the wiggle spectrum. The solid fuchsia line represents the mean amplitude in the wiggle-dominated part of the spectrum (shaded red region). WICKED flags spaxels by comparing this value to the standard deviation (blue dashed line). If data from both NRS detectors are available, as in this case, WICKED determines which part of the spectrum shows the most prominent wiggles and bases the flagging on that part of the spectrum.
    }
    \label{fig:FFT-example}
\end{figure*}

Where $F_{ratio}$ is the the Fourier ratio, and $f_{wiggles}$ represent the maximum frequency of the wiggles. The default Fourier ratio in {\sc WICKED} is $F_{ratio} = 3$ and we recommend to use always a value of $F_{ratio}\geq1.5$ (see Section~\ref{subsec:FT_test}). 
Figure \ref{fig:FFT-example} shows an example of how the wiggle detection is performed in {\sc WICKED} for the brightest spaxel of the F170LP cube of the A-star J1757132. Figure \ref{fig:FFT-example} shows the output of {\sc FindWiggles.plot\_wiggle\_FFT} built-in method in {\sc WICKED} developed to examine the Fourier spectrum and the Fourier ratio of a particular spaxel.  The output of {\sc plot\_wiggle\_FFT} consists of 3 panels, showing the spectrum of the spaxel, the wiggle spectrum and its Fourier spectrum. We show two of the three panels in Figure~\ref{fig:FFT-example}. Figure \ref{fig:FFT-example} left, shows the resulting wiggle spectrum (red, described in Section~\ref{subsec:wiggle_model}) and the masked regions (pink). The right panel shows the Fourier spectrum for the NRS1 (red) and NRS2 (dark red) part of the spectrum. The red shaded region shows the range of frequencies dominated by wiggles (identified during Step 1, see \S~\ref{subsec:wiggle_model}). The Fourier ratio for the brightest spaxel of the A-star J1757132, is $\sim 5.5$. This ratio is calculated as the mean amplitude of the NRS Fourier spectrum affected by wiggles (horizontal fuchsia line) divided by the 1-sigma value (mean amplitude + standard deviation) of the rest of the Fourier spectrum (blue horizontal dashed line).

{\sc WICKED} calculates the Fourier spectrum separately for both NRS detectors and uses the part of the spectrum where the Fourier ratio is largest to flag the spaxel. This approach is adopted because, in general, the spectrum in one NRS detector is more affected by wiggles than the other depending at which wavelengths the source is more compact. Using the spectrum from the detector less affected by wiggles would lower the overall mean amplitude at wiggle-dominated frequencies, reducing the ability to accurately flag the spaxel.

During the {\sc FindWiggles.get\_wiggly\_pixels} step, each  Fourier spectrum is calculated in parallel inside a search radius (default 10 pixels) which speeds up the process. Since the Fourier ratio for a given spaxel is only known after this step is complete, we developed a method called {\sc FindWiggles.define\_affected\_pixels} to set the Fourier ratio threshold. This function allows users to define a Fourier ratio threshold and view a map of flagged spaxels across the data cube, without the need to re-run {\sc FindWiggles.get\_wiggly\_pixels} each time the threshold is adjusted. Once an optimal threshold is defined using {\sc FindWiggles.define\_affected\_pixels}, the {\sc FitWiggles.fitwiggles} step is used to fit and remove the wiggles from each spaxel. The corrected spectra are then saved in a new data cube.

\section{Quantifying {\sc WICKED}'s Performance}
\label{sec:wicked_tests}

In this section we show the results of different tests designed to evaluate {\sc WICKED}'s ability to identify wiggles and recover the spectrum at different {\it S/N} ratios. We also compare {\sc WICKED}'s performance to the method by \citet{Perna2023}. The tests were performed using the spectrum of the A-star J1757132 and the M-star J15395077. The PSF aperture-extracted (3 pixel radius) spectra of these stars serve as known, wiggle-free references for our test. 
The A-star J1757132 was chosen for its high {\it S/N} ($\sim 500$) and prominent hydrogen lines, which we use to test the impact of the code on the line strength measurements. We created a wiggle model from the difference of the integrated spectrum of J1757132 and its brightest spaxel. In the upper left panel of Figure~\ref{fig:FFT_test_SNR} the integrated spectrum of the A-star J1757132 is shown in black, and the wiggle model in green. 

To simulate various scenarios, the wiggle model was manually added to the integrated spectrum of J1757132 along with Gaussian random noise to achieve different target {\it S/N} ratios. Using this method, we created datacubes for the integrated spectrum of J1757132 at {\it S/N} ratios of $\sim 500$, $400$, $300$, $200$, $150$, $100$, $50$, $15$, $10$ and $8$. For each datacube, we saved the noisy spectrum with wiggles, and the noisy spectrum without wiggles at different spaxel in the outer parts of the data cube. This allowed us to process the datacube using WICKED and the code by \citet{Perna2023} without modifying the codes. 
To account for Gaussian random noise variability, we created ten individual datacubes for each {\it S/N} ratio by adding different noise realizations, resulting in a total of 100 datacubes. Generating multiple cubes for each {\it S/N} ratio also allowed us to quantify systematic uncertainties during the wiggle removal by both {\sc WICKED} and the method by \citet{Perna2023}.

In Section~\ref{subsec:FT_test} we present the ability of our Fourier-ratio-based method to identify wiggles at different {\it S/N} ratios. In Section~\ref{subsec:WICKED_EW_test} we further test the stability of the shape of absorption lines and the continuum in the spectrum (S~\ref{subsec:WICKED_EW_test}). 
Finally, in Section~\ref{subsec:LOSV_WICKED} we evaluate how well {\sc WICKED} preserves the overall kinematics by looking at the integrated spectrum of the M-star J15395077. 

\subsection{Fourier ratio sensitivity to detect wiggles}
\label{subsec:FT_test}

\begin{figure*}

    \includegraphics[width=0.33\textwidth]{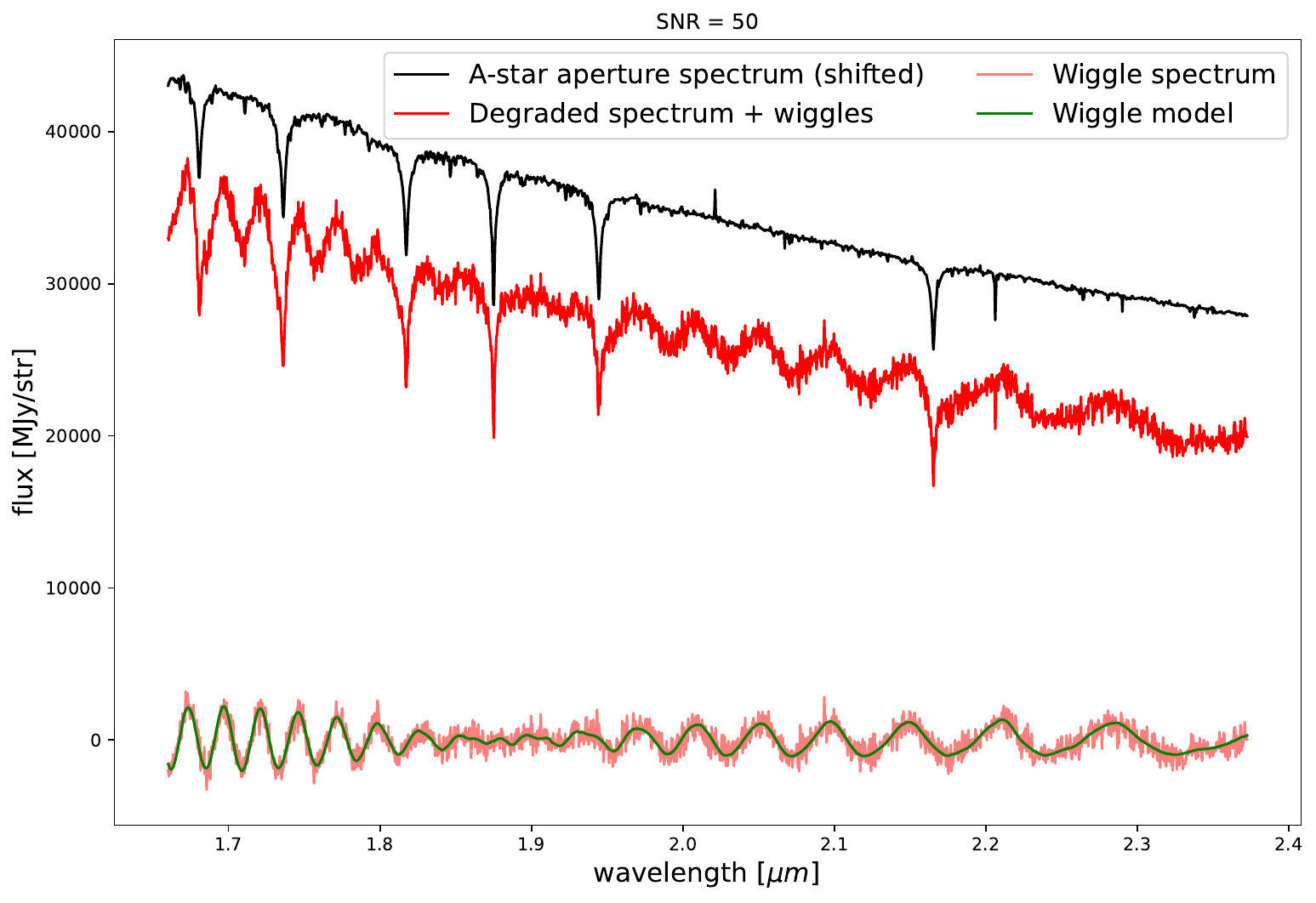}
    \includegraphics[width=0.33\textwidth]{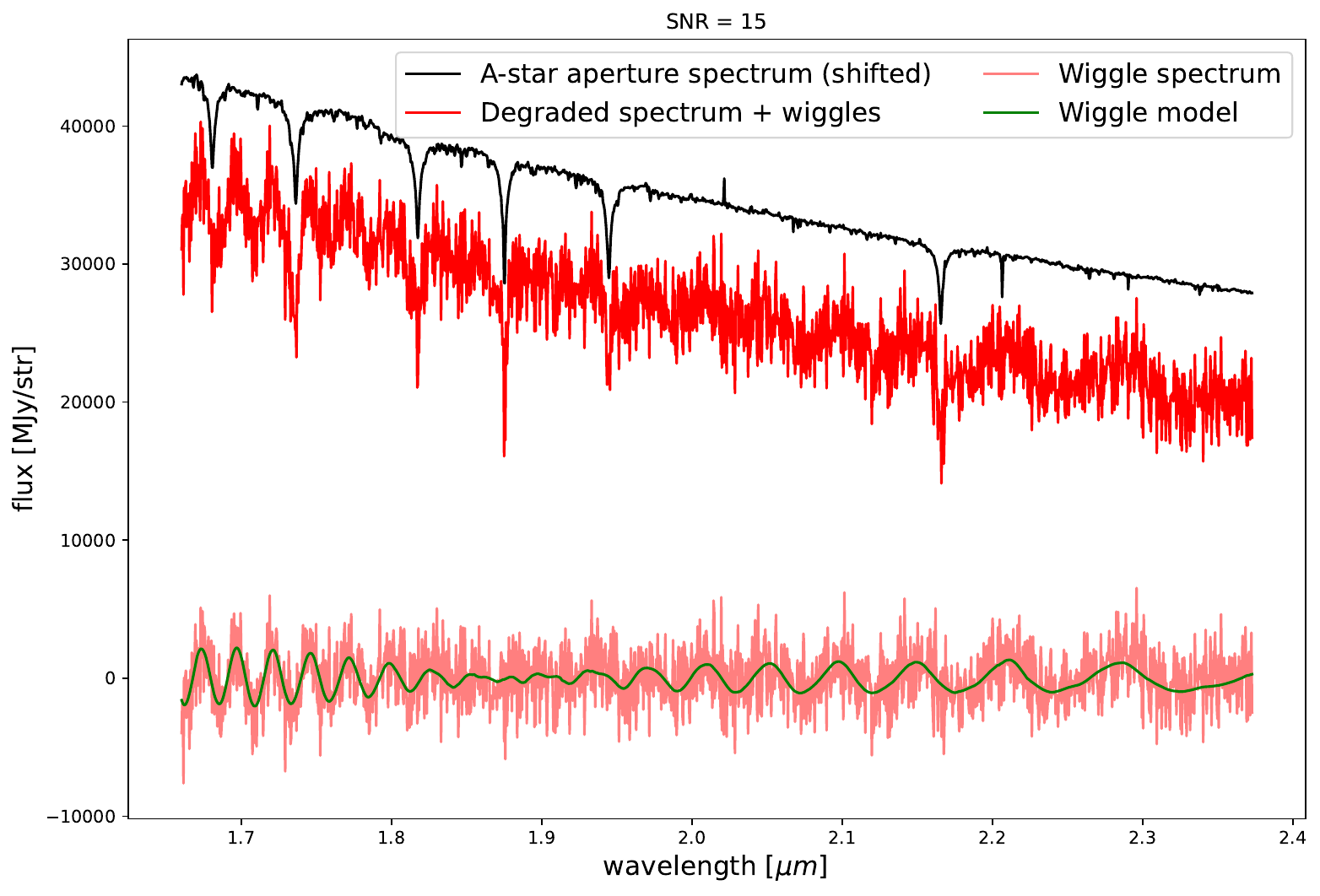}
    \includegraphics[width=0.33\textwidth]{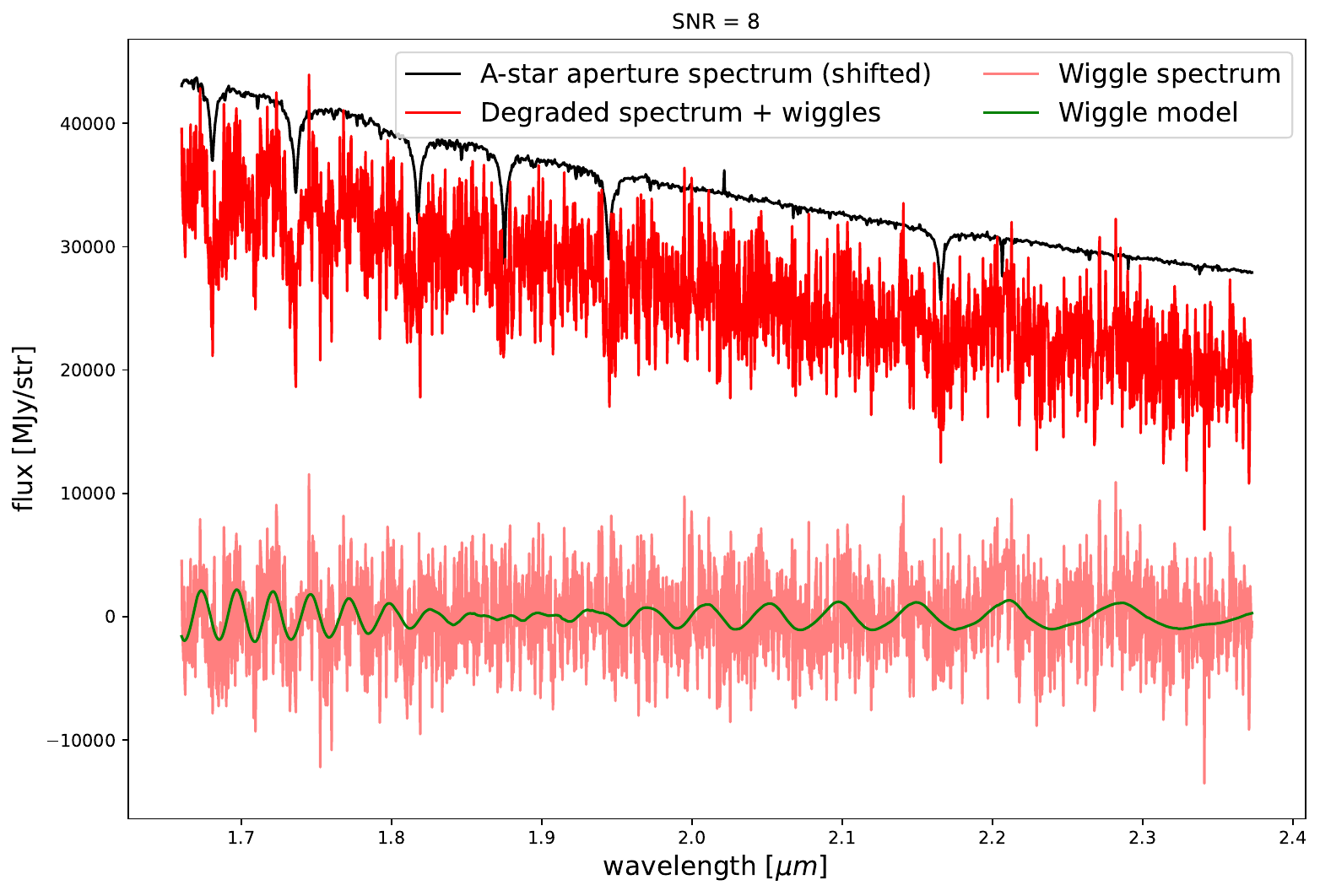}

    \includegraphics[width=0.33\textwidth]{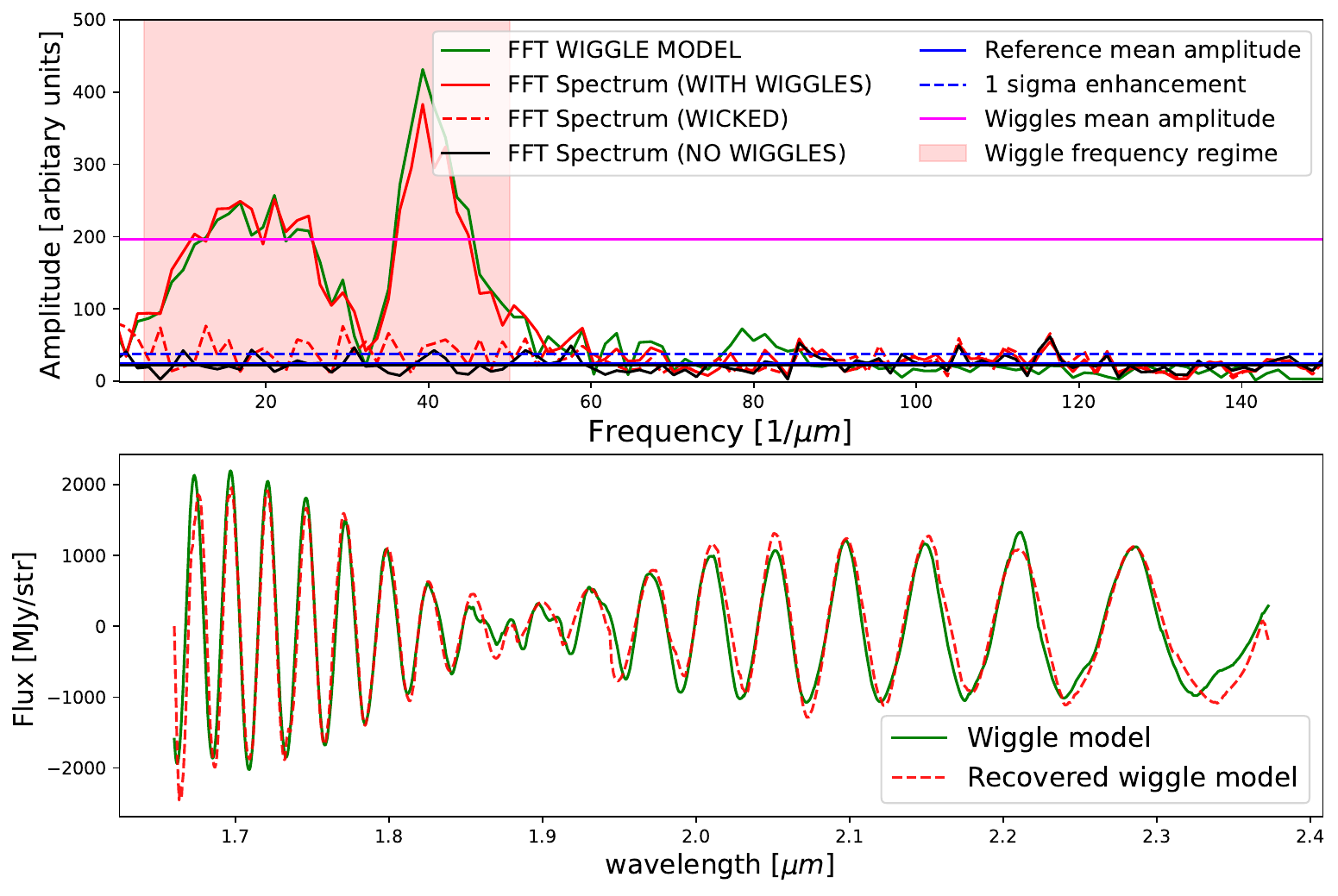}
    \includegraphics[width=0.33\textwidth]{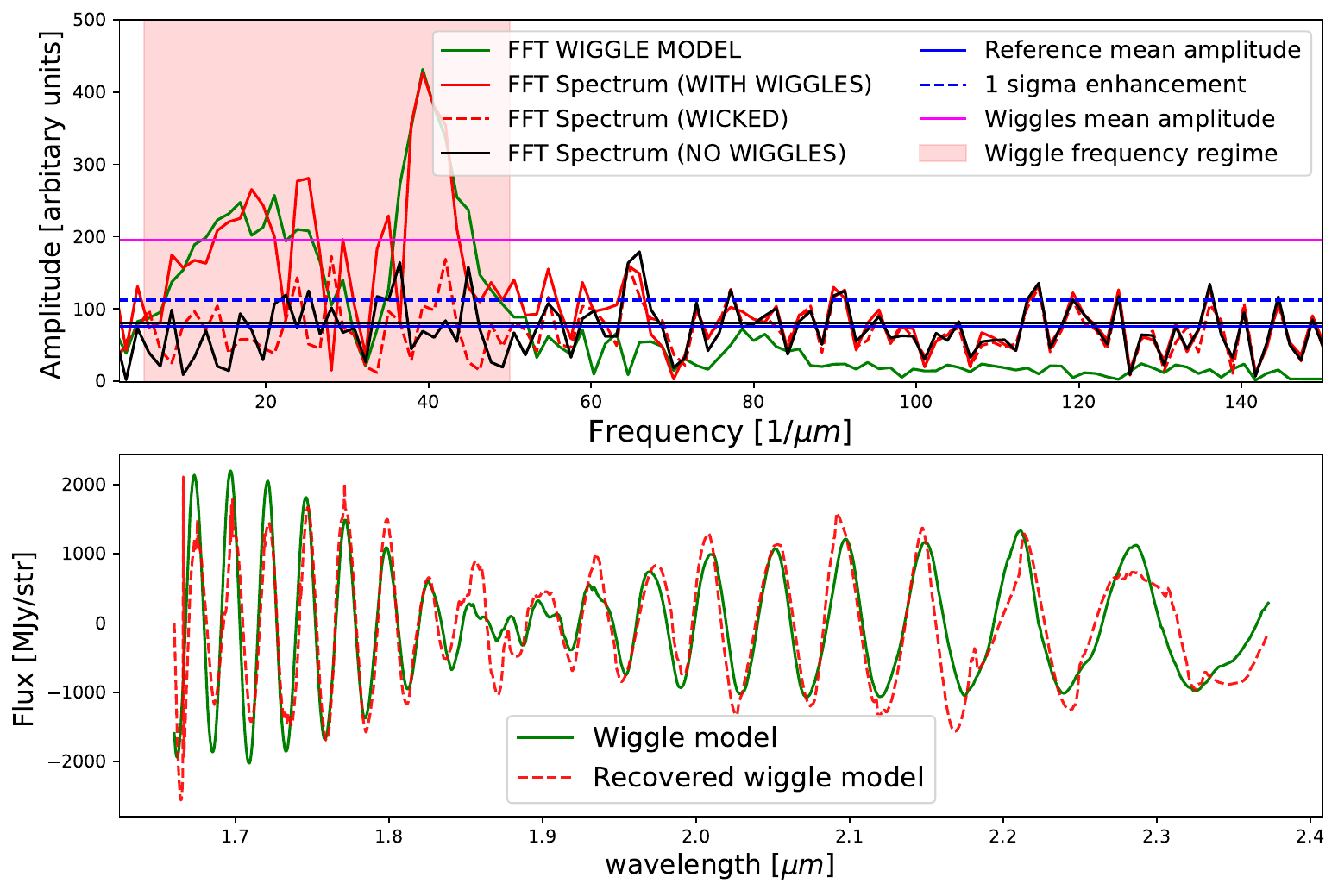}
    \includegraphics[width=0.33\textwidth]{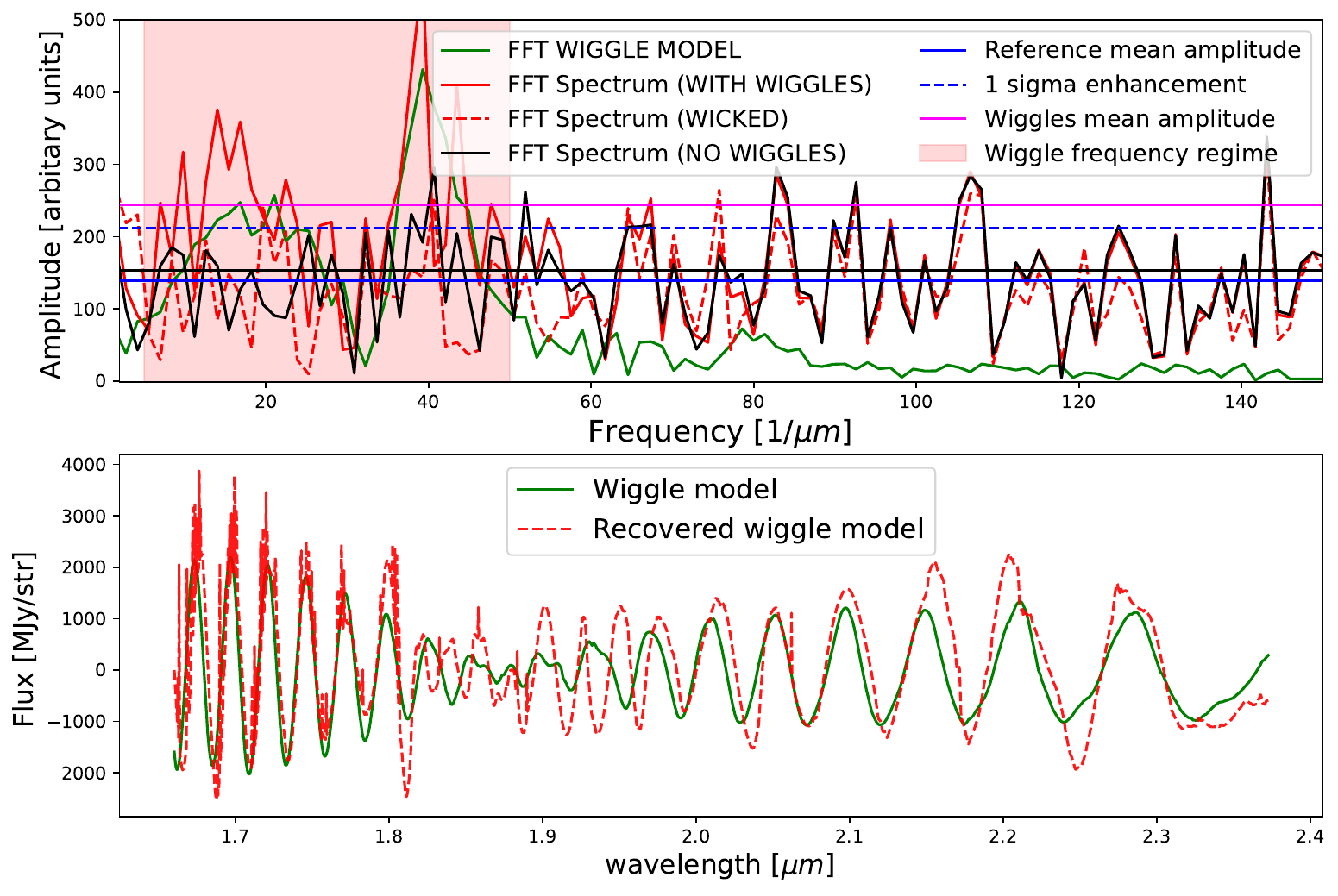}
\caption{Comparison of the spectrum of the A-star J1757132 (top, black) and the degraded spectrum (top, red) with the added wiggle model (green) at {\it S/N} ratios of $50$, $15$ and $8$. 
{\it Middle:} the Fourier spectra of i) the degraded spectrum (solid, red), ii) the wiggle model (green), iii) the degraded spectrum without wiggles (black), and iv) the data corrected with {\sc WICKED} (dashed, red). 
The horizontal lines mark the mean amplitude of the Fourier spectrum at frequencies dominated by wiggles and at larger frequencies. The Fourier ratio can effectively distinguish wiggles from noise down to a {\it S/N} ratio of $\sim 8$. {\it Bottom:} comparison of the input wiggle model (green) versus the recovered wiggle spectrum for the data cleaned with {\sc WICKED}.}
\label{fig:FFT_test_SNR}

\end{figure*}

To get a sense of up to which {\it S/N} the Fourier ratio method used in {\sc WICKED} can reliably detect wiggles in a spectrum, we examined the Fourier spectrum of J1757132 at different {\it S/N} values and compared the Fourier ratio for spectra with and without the wiggle model. 
In Figure~\ref{fig:FFT_test_SNR}, we present the Fourier spectra for our three lowest {\it S/N} cubes ($50$, $15$, and $8$) to illustrate how this method separates wiggles from noise. The original aperture spectrum of J1757132 ({\it S/N} $\sim 500$) is shown in the top panel in black, and the degraded spectra in red. In the top panel of Figure~\ref{fig:FFT_test_SNR} we also show the original wiggle model (green) added to the data, and the version with added noise (red). The wiggle spectrum (top, red) is obtained by subtracting the best-fit model from the degraded spectrum (with added noise) and represents what {\sc WICKED} will try to fit, as described in Section~\ref{subsec:wiggle_model}. However, as the {\it S/N} decreases, the wiggles begin to blend with the noise, making them harder to detect by visual inspection alone. For spectra with {\it S/N} $\leq 15$, distinguishing wiggles from noise becomes unreliable without a mathematical approach like the Fourier ratio.

The middle panel of Figure~\ref{fig:FFT_test_SNR} shows the Fourier spectrum for each spectrum at a given {\it S/N} (solid, red), for the degraded spectrum without adding the wiggle model (black), and for the spectrum corrected with {\sc WICKED} (dashed, red). The Fourier spectrum of the wiggle model (green) is also shown in the middle panel of Figure~\ref{fig:FFT_test_SNR}. The Fourier spectrum of the wiggle model has two large peaks at frequencies $f_{\lambda} \approx 20, 40$ [$\mu m^{-1}$], but it has hardly any feature at frequencies $f_{\lambda} > 60$ [$\mu m^{-1}$]. In contrast, the Fourier spectrum for the degraded spectrum without the added wiggles (black) regardless of the {\it S/N}, shows similar levels of features at all frequencies and lacks the two peaks at $f_{\lambda} \approx 20, 40$ [$\mu m^{-1}$] seen in the wiggle model Fourier spectrum. The Fourier spectrum for the degraded spectrum with added wiggles (red) shows the spectral features at $f_{\lambda} > 60$ [$\mu m^{-1}$] almost identical to the spectrum without the added wiggles, and the two prominent peaks from the wiggle model at $f_{\lambda} \approx 20, 40$ [$\mu m^{-1}$]. Finally, the Fourier spectrum of the data cleaned with {\sc WICKED} closely matches the reference Fourier spectrum of the data without wiggles, showing that the two prominent peaks have been largely removed, and at a {\it S/N} of 8 we still remove $\sim$50\% of the wiggle signal.  

The horizontal lines in Figure~\ref{fig:FFT_test_SNR} are similar to the ones in Figure~\ref{fig:FFT-example}, but since here we compare three Fourier spectra at once we have changed the colors. The fuchsia line in Figure~\ref{fig:FFT_test_SNR} marks the mean amplitude for the degraded spectrum with added wiggles, and the black line for the degraded spectrum without wiggles (in the wiggle regime). These two values are never the same at any {\sc S/N}, while the blue line is almost identical to the mean amplitude (solid blue line) at frequencies outside the wiggle regime of the spectrum with the added wiggles. This shows that the wiggles and the spectral features are in completely separate parts of the Fourier spectrum. 
When we look at the Fourier ratio, used to flag spaxels in {\sc WICKED}, we see a clear tendency; at higher {\sc S/N} the Fourier ratios are higher ($\sim 14$ at {\it S/N} of $50$) and $\sim 1.5$ at {\it S/N} $8$. The Fourier ratio changes because the 1-$\sigma$ benchmark (blue dashed-line) for the part of spectrum outside frequencies dominated by wiggles increases at lower {\sc S/N}, while the mean amplitude for the wiggle regime (fuchsia line) remains almost the same. 

The bottom panel of Figure~\ref{fig:FFT_test_SNR} shows the recovered wiggle spectrum with {\sc WICKED} and compares it to the true wiggle model. At a {\it S/N} of $100$ both are mostly identical, having a mean difference of only $1.7 \pm 1.9 \: \%$ and a maximum difference of $\sim 7 \:\%$ for the wiggles at $\sim 1.9$ $\mu m$. At a {\it S/N} of $15$ we can see larger difference not only in the lower amplitude wiggles around $\sim 1.9$ $\mu m$ but also in the more prominent wiggles around $1.3$ $\mu m$. However, the differences are still $3.4 \pm 2.8 \: \%$ on average with a maximum value of $\sim 12 \: \%$ around $\sim 2.15$ $\mu m$. Interestingly, this is close to an absorption line (LINE 3 in Figure~\ref{fig:EW_test}) that is masked during the cleaning with {\sc WICKED}, which could explain the larger difference. Finally, at a {\it S/N} of $8$ we observe the largest difference in the recovered wiggle spectrum, with an average difference between the {\sc WICKED}-clean and degraded aperture spectrum of $5.5\pm4.5 \: \%$, only $\sim 500$ [MJy/sr], which is about 5$\times$ smaller than the average noise ($\sim 2700$ [MJy/sr]). 

In summary, the Fourier transform is quite reliable at distinguishing wiggles in the spectrum versus Gaussian random noise. The Fourier ratios are more than $\geq 3 \: \sigma$ larger than the amplitude of the spectral features for spectra with a {\it S/N} $\geq 15$ in the presence of large wiggles as simulated by our wiggle model. The Fourier ratio decreases quickly, becoming only $\sim 1.5 \: \sigma$ at a {\it S/N} of $8$. The recovered wiggle spectrum after cleaning the spectrum with added noise in {\sc WICKED}, shows insignificant differences at high {\it S/N}, with more than 80\% of the wiggles signal removed at {\it S/N} of 50. At lower {\it S/N} the mean differences increase, and there are also more often larger differences between the recovered and the original wiggle model. However, as mentioned above this difference are still in average smaller than the noise and with most of the wiggles signal removed from the data ($\sim$50\%). Based on this, we recommend using {\sc WICKED} with a Fourier ratio $F_{ratio} \geq 1.5$.
We cannot directly compare {\sc WICKED}'s ability to flag spaxels with wiggles to the code developed by \citet{Perna2023}, as their method relied solely on a {\it S/N} threshold based on the brightness of the brightest spaxel which leads to many spaxels being incorrectly flagged or missed entirely.

\subsection{Preservation of continuum and lines shape}
\label{subsec:WICKED_EW_test}

To evaluate how well {\sc WICKED} preserves the shape of the absorption lines in the spectrum, we study the equivalent width of three absorption lines in the spectrum of J1757132 at various {\it S/N} ratios. There are no gas emission lines in the spectrum of J1757132, but there should be no practical difference on how {\sc WICKED} deals with emission or absorption lines. We also evaluate how {\sc WICKED} preserves the overall spectral shape by comparing the mean difference of the spectrum at different {\it S/N} ratios with respect to the aperture spectrum of J1757132. We also do this for the spectra without the added wiggles, to simulate the impact of wrongly flagged spaxels cleaned in {\sc WICKED}, and its impact in its spectral shape. We compare all our results with the cubes cleaned using the method by \citet{Perna2023}.

\begin{figure*}
\includegraphics[width=1.\textwidth]{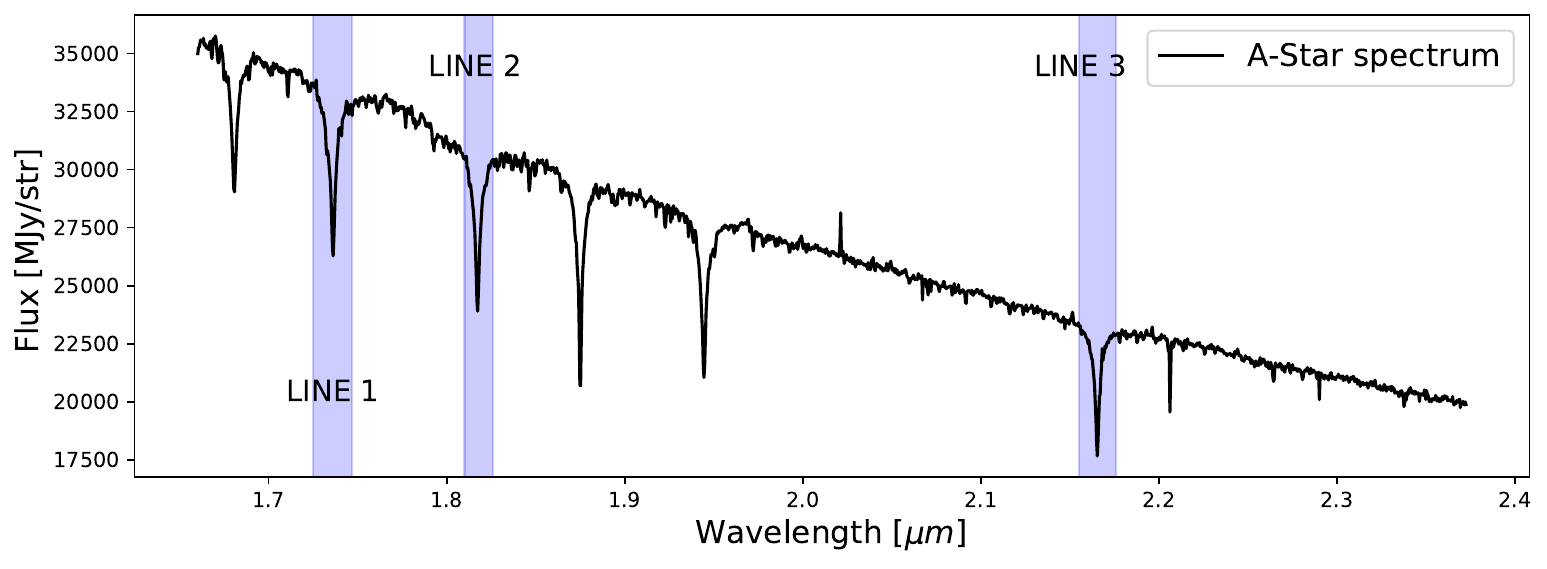}
\includegraphics[width=1.\textwidth]{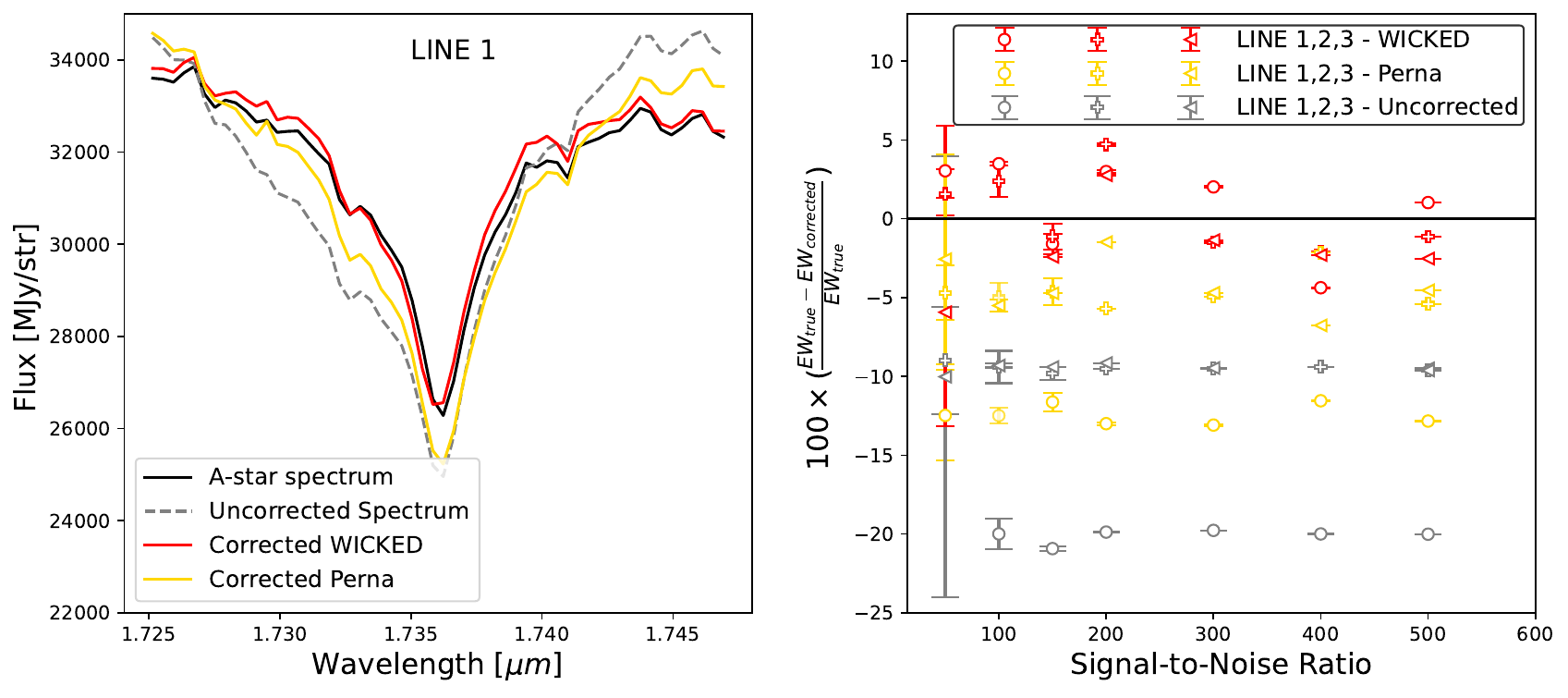}

\caption{Comparison of recovered equivalent width for two absorption lines present in the spectrum of the A-star (solid black line) J1757132. The bottom left panel shows the profile of the LINE 1 for the original spectrum (solid black), the spectrum with wiggles (dashed) and the spectra corrected using WICKED (red) and with the code by \citet{Perna2023} (yellow). The line profile and equivalent width are better recovered with WICKED than with the code by \citet{Perna2023}, with the corrected spectrum having $< 5\%$ difference with the ``true'' value derived from the spectrum with no wiggles. For the uncorrected spectrum (gray) we see EW differences exceeding the typical error (~4\%) in spectra with {\it S/N} as low as 50.}
\label{fig:EW_test} 
\end{figure*}

The equivalent width (EW) for each line was determined by multiplying the full width at half maximum (FWHM) by the flux at the minimum of the absorption line. The errors in the EW were calculated using the 50 Monte Carlo simulations based on the error array. The results of the test for the EW are shown in Figure~\ref{fig:EW_test}. The top panel of Figure~\ref{fig:EW_test} shows the aperture spectrum of J1757132 and the three absorption lines used for the equivalent width comparison marked in blue. We compare the EW for three different lines since they are affected differently by wiggles because the amplitude and frequency of the wiggles change with wavelength. The bottom left panel shows the profile for one of these lines, ``LINE 1''. The solid black line is the line profile for the aperture spectrum ({\it S/N} of $500$), the dashed gray line for the uncorrected spectrum (the output of the JWST pipeline), the red line for the spectrum cleaned with {\sc WICKED}, and finally in yellow for the spectrum cleaned with the method by \cite{Perna2023}. The right panel shows the percentile difference for each line for the different data cubes at different {\it S/N} compared to the ``true'' equivalent width. The true equivalent width is defined as the value obtained from the degraded spectrum at that {\it S/N} without added wiggles. 

The differences in EW are quite similar across all different {\it S/N} ratios, and they are mostly dominated by the amplitude of the wiggle model. For the uncorrected spectrum (gray symbols), LINE 1 that is in the region of the spectrum with large amplitude wiggles, has a difference of $\sim20 \: \%$, while the rest of the lines show a difference of $\sim 10\%$. Based on this test we expect EW biasses larger than typical error ($\sim$4\%) down to spectra with a {\it S/N} of 50. For the spectra cleaned using {\sc WICKED} (red symbols), we see similar differences in EW regardless of the {\it S/N} and the absorption line with a mean difference of only $3.5 \: \%$. This difference is $\sim 4\times$ smaller than the average difference for the uncorrected spectra ($14\:\%$), and more than $2 \times$ smaller than the average difference for the spectra cleaned with the method of \citet{Perna2023} ($7.2\:\%$). This shows the ability of {\sc WICKED} to remove wiggles of different amplitudes and frequencies across the whole wavelength range with the same quality. This is not the case for the spectra cleaned with the method by \citet{Perna2023}, where LINE 1 has almost $2\times$ larger difference in EW as the other two lines. LINE 1 is located near the low-wavelength edge of the spectrum, while the other two lines are in the middle. We observe this discrepancy in the ability to remove wiggles at the edges of the spectrum in the method of \citet{Perna2023} across all our tests in this work (see, for example, Figure~\ref{fig:Comparison_extinsion}). 

\subsubsection{Mean flux difference}
\label{subsubsec:mean_flux_diff}
To quantify how the overall shape of the continuum is preserved during the cleaning with {\sc WICKED}, we calculate the mean flux difference of the spectrum with added wiggles at various {\it S/N} versus the aperture spectrum. This helps to quantify the overall disagreement between the ``true'' (the aperture spectrum of J1757132) spectrum and a corrected spectrum processed with either {\sc WICKED} or the method by \citet{Perna2023}. The uncertainties are calculated from the standard deviation of the mean flux difference for the 10 corrected cubes at this particular {\it S/N}. As mentioned above we created 10 cubes at each {\it S/N} to help constrain the systematic errors introduced by cleaning the spectrum. Additionally, miss classification of spaxels across the data cube is a possibility, thus we also test the impact in the spectrum in the case of a spaxel with a ``clean'' spectrum (i.e without wiggles) that was wrongly flagged and cleaned with {\sc WICKED} and the code by \citet{Perna2023}.  

\begin{figure*}
\includegraphics[width=1\textwidth]{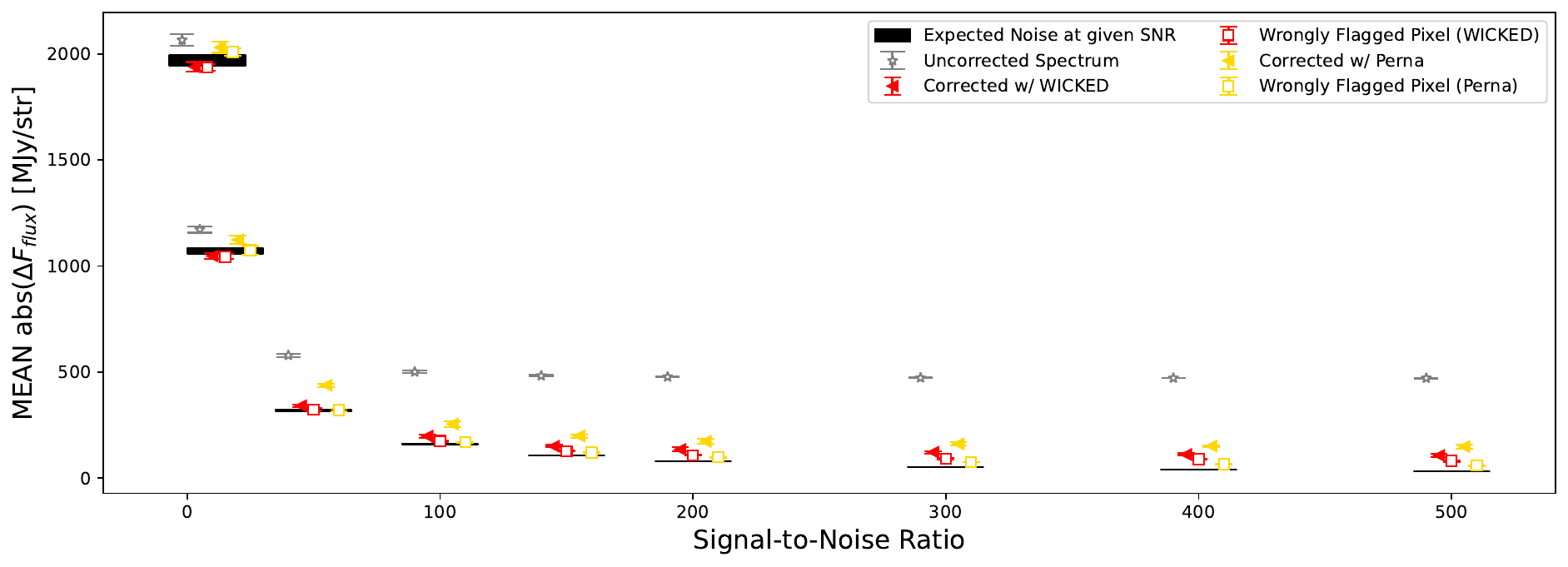}
\caption{Comparison of the mean difference flux respect to the aperture spectrum of the A-star J175713. The black rectangles represent the minimum expected the flux difference at a given {\it S/N} ratio. The spectra corrected using {\sc WICKED} (solid red) shows the smallest mean flux difference compared to the uncorrected spectrum (gray) and the spectrum clean using the method by \citet{Perna2023} (solid yellow) at all {\it S/N}.  }
\label{fig:MeanDiff_vs_SNR} 
\end{figure*}

The results are shown in Figure~\ref{fig:MeanDiff_vs_SNR}. The mean flux difference between the spectrum without added wiggles at that particular {\it S/N} and the aperture spectrum are shown as black rectangles in Figure~\ref{fig:MeanDiff_vs_SNR}, with the height representing the 1~$\sigma$ uncertainty. These black rectangles serve as a benchmark for comparing how well the overall shape of the spectrum is preserved. In Figure~\ref{fig:MeanDiff_vs_SNR} we can see that the mean flux difference for the uncorrected spectrum (shown in gray stars) is dominated by the amplitude of the wiggles at high {\it S/N}, while at {\it S/N} $\leq 50$ by the Gaussian random noise of the spectrum. At low {\it S/N} ratios some of the differences introduced in the spectrum by the wiggles get lost in the noise of the spectrum. 

For the spectrum with wiggles (solid symbols) in Figure~\ref{fig:MeanDiff_vs_SNR}, the spectra cleaned with {\sc WICKED} (red squares) show the smallest mean flux difference across all {\it S/N} when compared to the uncorrected spectra and the ones cleaned using the method from \citet{Perna2023} (yellow squares). On average the spectrum cleaned with {\sc WICKED} has a $\sim 4\times$ smaller mean flux difference compared to the uncorrected spectrum, and $\sim 1.4\times$ than the one cleaned using the method by \citet{Perna2023} down to a {\it S/N} ratio of $50$. This means that if a spaxel in the data cube is correctly flagged (to have wiggles) the resulting spectrum will always be better than the output from the JWST pipeline, or by cleaning it with the method by \citet{Perna2023}.
 
The spectra without wiggles that were still cleaned using {\sc WICKED} and the method by \citet{Perna2023} are shown in open squares in Figure~\ref{fig:MeanDiff_vs_SNR}. At high {\it S/N} both codes ``impair'' the spectrum by trying to correct for wiggles when they are not present, although the differences are very small. At {\it S/N} of $500$ and $400$ the spectrum cleaned using {\sc WICKED} the difference is only $\leq 50$ [MJy/str], which is of the same order as the average error of the spectrum ($\sim 54$ [MJy/sr]). This difference becomes $\sim 25$ [MJy/str] at a {\it S/N} of $200$, which is $4\times$ smaller than the average error. At smaller {\it S/N} the difference continues to decrease until it becomes comparable to the uncertainties  of the benchmark rectangles. The mean flux differences are close to half smaller for the spectrum cleaned using the method by \citet{Perna2023} compared to {\sc WICKED} at {\it S/N} $\gtrapprox200$. However, the difference between the two methods becomes comparable at about {\it S/N} of $150$.  This difference between {\sc WICKED} and the method by \citet{Perna2023}, and the fact that the mean flux difference for {\sc WICKED} seems to be slightly lower than the benchmark rectangles at very low {\it S/N}, could indicate that {\sc WICKED} is more prone to over-fitting. However, as mentioned above, we emphasize that the differences are very small.
Finally, as shown in Subsection~\S~\ref{fig:FFT_test_SNR} {\sc WICKED} is quite reliable at flagging spaxels affected by wiggles down to a {\it S/N} $\sim 8$, while the method by \citet{Perna2023} relies only on a total flux comparison with the brightest spaxel, which leads to more spaxels flagged incorrectly. Thus, it is unprovable for {\sc WICKED} to incorrectly flag a spaxel at high {\it S/N} ratios, and consequently affecting the quality of the spectrum. 

In our test we have observe that a high {\it S/N} spectrum can be misclassified and significantly affected if there is a large mismatch between the best fit-model and the data. We recommend the users to never run {\sc WICKED} blindly, and always inspect for spaxels that could be incorrectly flagged. These spaxels can be excluded from the fitting using the keyword {\sc define\_affected\_pixels.exclude\_pixels} (Please see the example {\sc Jupyter Nootebook} in the {\sc Github} repository for details).

\subsubsection{Changes in the spectra across the field-of-view}
\label{subsubsec:change_in_continuum}
Spatial variations in the spectra across the data cube, such as those caused by differential dust obscuration or contributions from different sources, can lead to mismatches between the two stellar templates and the spectrum at a given spaxel. In such cases, the best-fit model may rely more heavily on the power-law and second-degree polynomial to account for these differences.

We simulated this scenario by creating a ``dust-obscured'' spectrum of the aperture spectrum of J1757132. We used the infrared-optical extinction law from \citet{Cardelli1989} to add dust component with a value of A$_{V}=20$. The obscured spectrum was saved in the spaxel X$_{\rm spaxel} = 16$, Y$_{\rm spaxel} = 16$ and then cleaned using {\sc WICKED} and the method by \citet{Perna2023}, similarly as done for the previous cubes (S~\ref{sec:wicked_tests}).

\begin{figure*}
\includegraphics[width=0.5\textwidth]{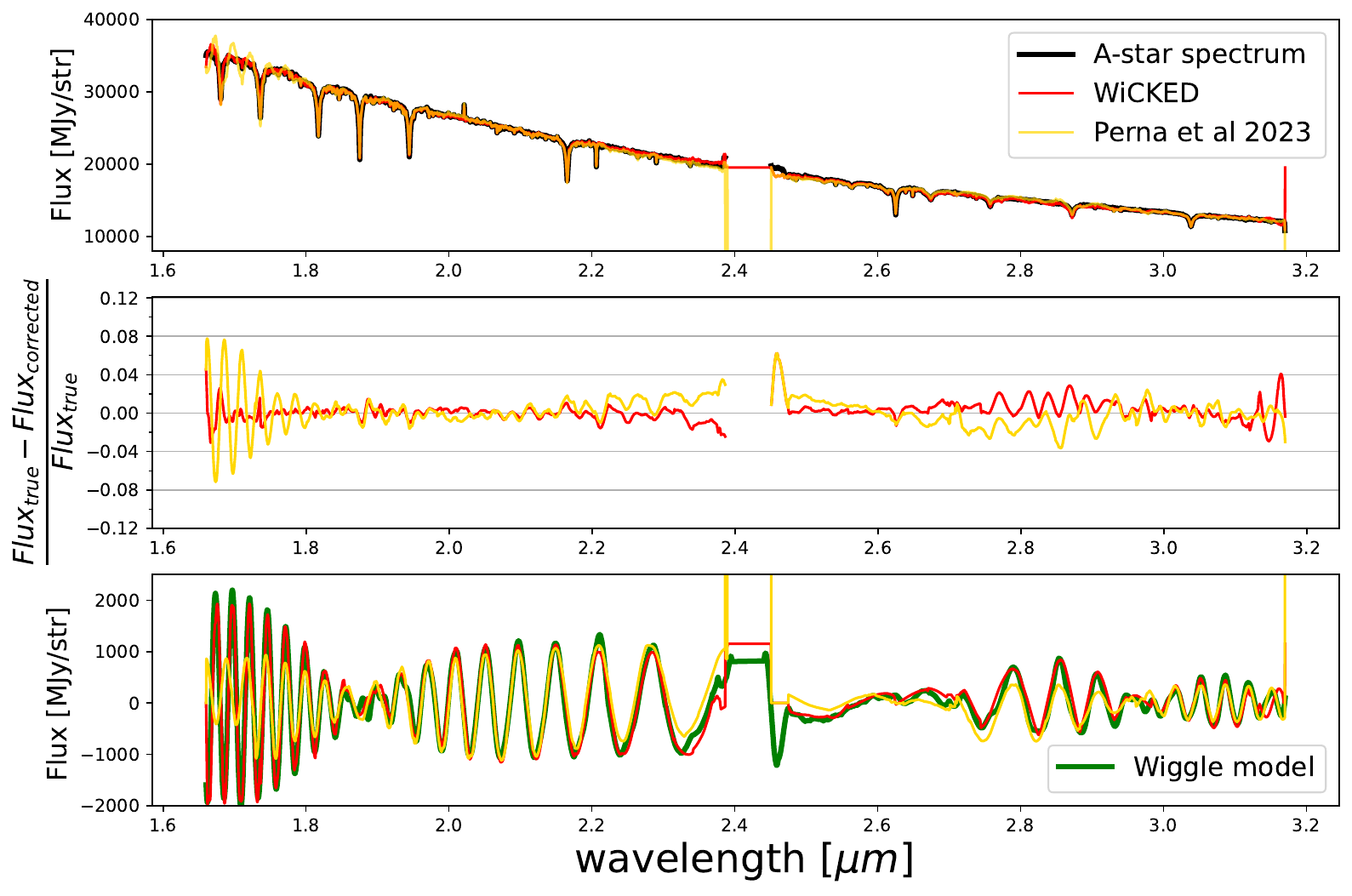}
    \includegraphics[width=0.5\textwidth]{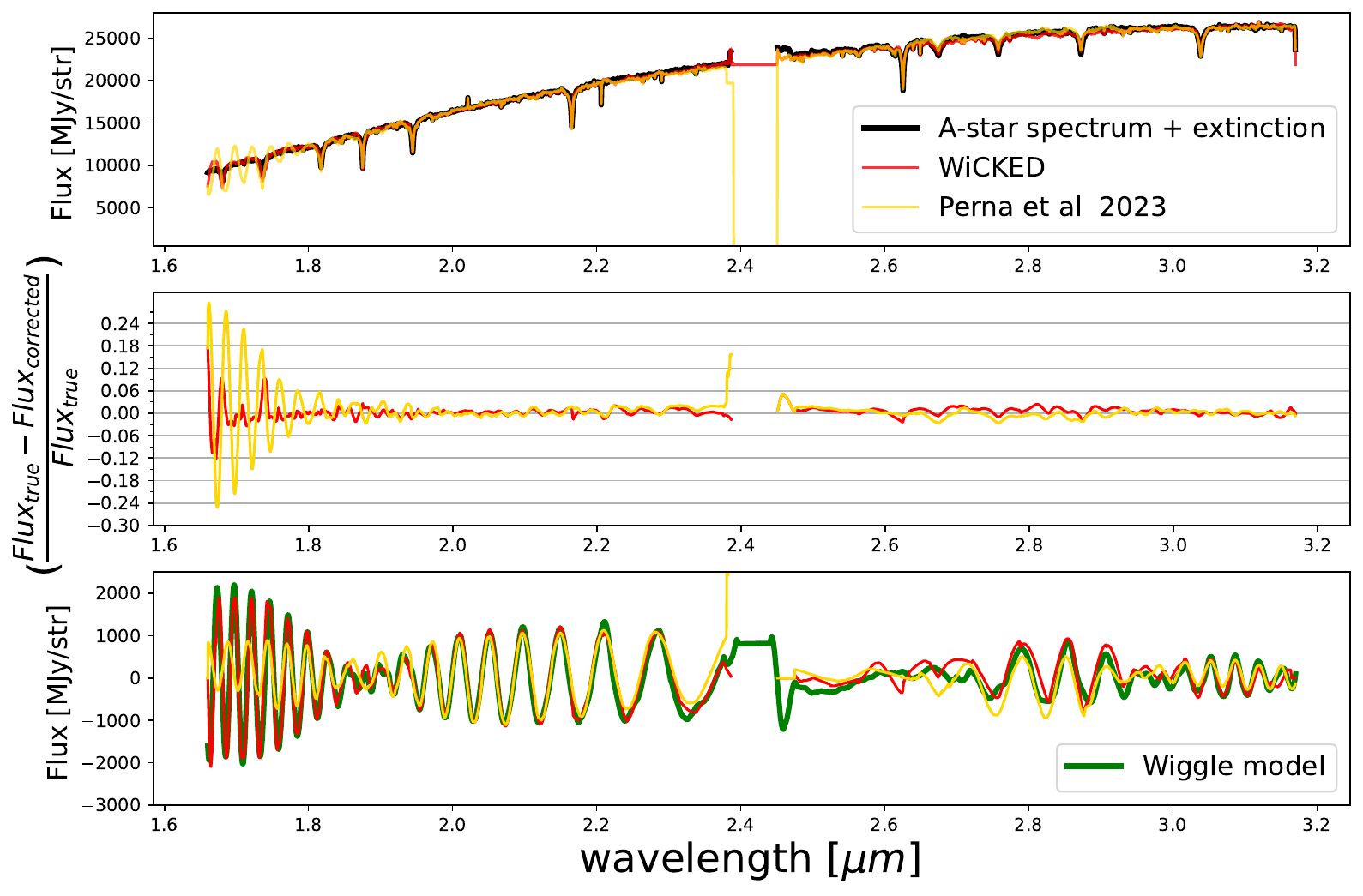}
\caption{Comparison between {\sc WICKED} and the code by \citet{Perna2023} at correcting for wiggles. The left panel shows the performance of both codes at recovering the wiggles model (shown in black in the bottom panel) added to the spectrum of the A-star J1757132 (solid black line). In the right panel we compare the performance of both codes at recovering the wiggles when the shape of the continuum is different from the integrated spectrum. We modified the spectrum by adding dust extinction with a value of $A_{V} =20$, using the Infrared-Optical relation by \citet{Cardelli1989}. As shown {\sc WICKED} is better at recovering the wiggle then the code by \citet{Perna2023}, specially when the integrated spectrum is not a perfect model for the spectrum.}
\label{fig:Comparison_extinsion}
\end{figure*}

The resulting obscured spectrum is shown in black in the left panel of Figure~\ref{fig:Comparison_extinsion}. Figure~\ref{fig:Comparison_extinsion} also shows the clean spectrum using {\sc WICKED} (red) and the code by \citet{Perna2023} (yellow). We also show the clean spectrum without the added dust obscuration to have a side-by-side comparison of their differences. The middle panel shows the relative flux difference between the clean and the integrated spectrum. Colors are the same for all the panels. The bottom  shows the recovered wiggle spectrum by {\sc WICKED} and the code by \citet{Perna2023} and the original wiggle model (green). 

As expected from the results of section~\ref{subsubsec:mean_flux_diff}, the spectrum cleaned with {\sc WICKED} has lower relative differences with respect to the one done with the method by \citet{Perna2023} for the spectrum without dust obscuration. The spectrum cleaned with {\sc WICKED} shows relative flux differences around $0$ with some small section in the spectrum with differences of $\sim 4 \:\%$. However, the spectrum cleaned with the method by \citet{Perna2023} have a large portion in the spectrum between $1.7-1.8$ $\mu m$ with differences of $\sim8 \: \%$. The situation is worsened for the obscured spectrum, where 
the data cleaned with the method by \citet{Perna2023} shows about $3\times$ larger differences in that section, with relative differences of $\sim 25\:\%$. For the spectrum clean with {\sc WICKED} however, the differences are relatively the same as for the un-obscured data with no large portion of the spectrum with large differences, only having a maximum difference of $\sim 6\:\%$ around an absorption line in $\sim 1.8$ $\mu m$. The spectrum corrected with the code by \citet{Perna2023} also shows a small ``dip'' in the continuum for the un-obscured spectrum in the $2.7-2.9$ $\mu m$ region.  These results support the use of a power-law plus a simple polynomial (second-degree) to model mismatches between the spectrum and the integrated templates, over the high degree polynomial used in \citet{Perna2023}. 

\subsection{Enabling Line-of-sight velocity measurements with WICKED}
\label{subsec:LOSV_WICKED}

The presence of wiggles can significantly distort line shapes (see bottom-left panel of Fig.~\ref{fig:EW_test}), impacting the ability to obtain reliable kinematic fits for NIRSpec at a single-spaxel level. In this section we show how {\sc WICKED} can effectively subtract these artifacts to obtain reliable kinematics that otherwise could not be possible. For this we used the F170LP NIRSpec data cube of the M-star J15395077 rather than the previously used A-star J1757132. A-stars lack the Carbon Monoxide (CO) bandhead at $2.30 \: \mu m$ series commonly used for stellar kinematics, while they are the dominant stellar feature in M-stars. The spectrum of 2MASS J15395077 was reduced in the same way as the data of J1757132 and corrected for wiggles using both {\sc WICKED} and the code by \citet{Perna2023}. We obtain the line-of-sight velocities (LOSV) and velocity dispersion ($\sigma$) for the brightest spaxel of the F170LP cube J15395077, for three different cubes; first an uncorrected cube, second for the cleaned cube using {\sc WICKED}, and the last cube cleaned using the method by \citet{Perna2023}. The LOSV and $\sigma$ were extracted using the {\sc Python} implementation of the penalized spaxel fitting routine, {\sc pPXF} \citep{Cappellari2017} with a set of synthetic high-resolution stellar templates from the Phoenix library \citep{phoenixlib}. A fifth-degree additive polynomial was used to model the continuum differences between the spectrum and the templates. Uncertainties for the LOSV and $\sigma$ were estimated using a bootstrap method.  We compare the resulting LOSV and $\sigma$ against an aperture-extracted spectrum with a $0.15\arcsec$ (3-pixel radius), close to the PSF value. This effectively reduces the sinusoidal modulations since they are spatially correlated and thus tend to cancel out.
Since the quality of the {\sc pPXF} fit is a function of the {\it S/N} of the data, we degraded the aperture spectrum of J15395077 from its native {\it S/N} $\approx150$ to a {\it S/N} of $41$ to match the {\it S/N} of the brightest spaxel of J15395077. 

\begin{figure*}
    \includegraphics[width=0.5\textwidth]{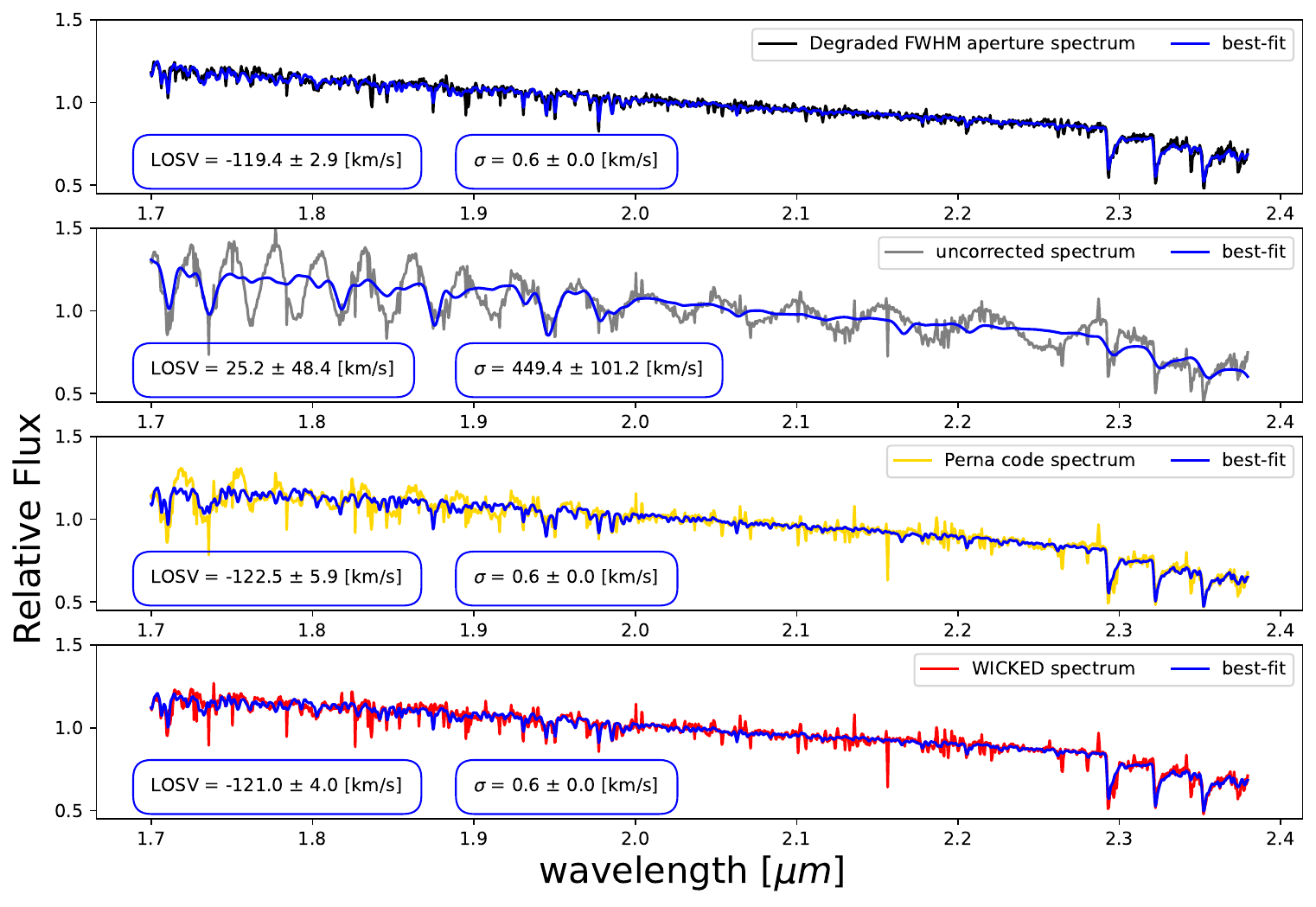}
    \includegraphics[width=0.5\textwidth]{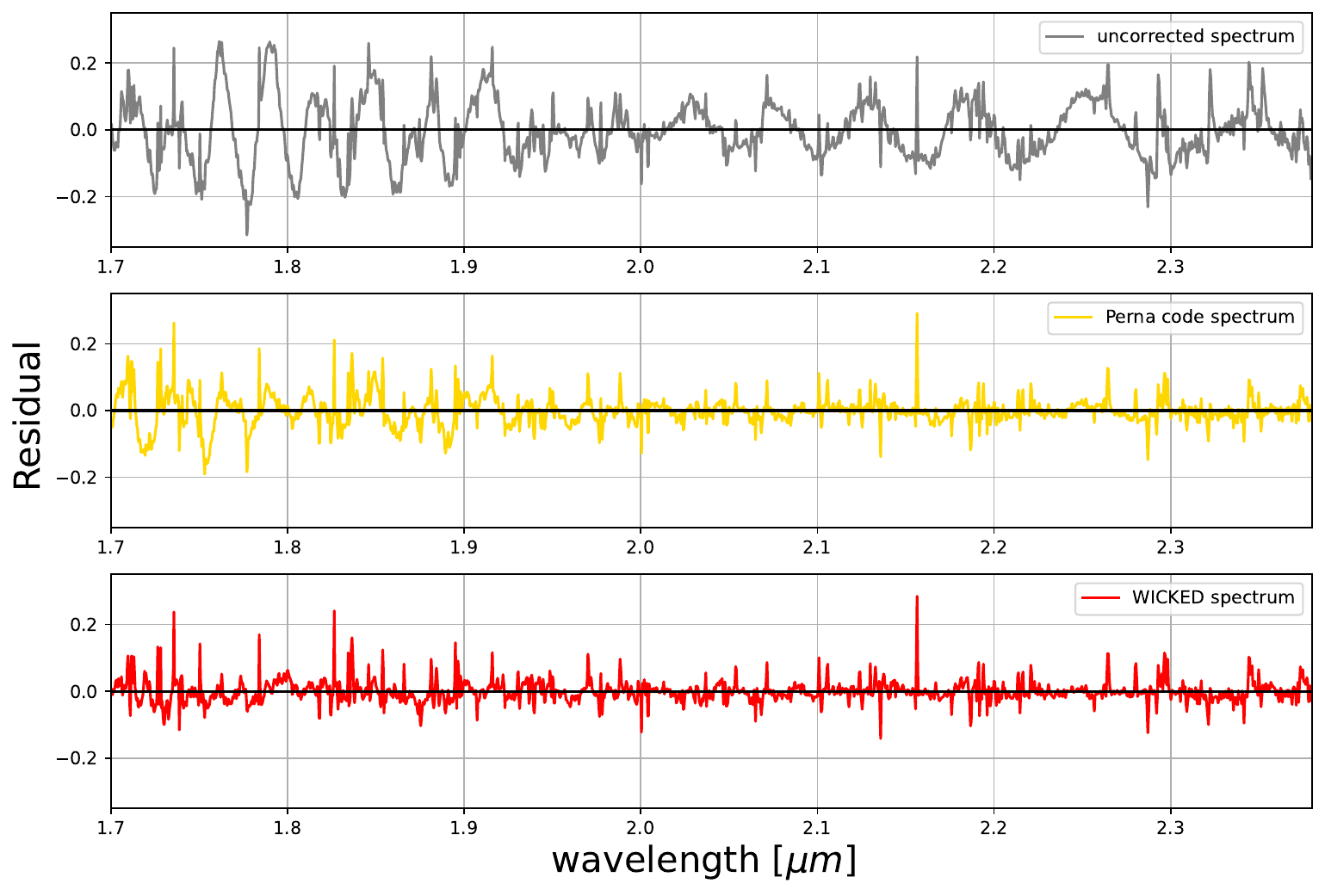}
\caption{\textit{Left:} {\sc pPXF} fits (blue) for the integrated spectrum (black) of the M-giant star J15395077, uncorrected brightest spaxel spectrum (gray), spectrum corrected with the code by \citet{Perna2023} (yellow) and corrected using {\sc WICKED} (red). The FWHM aperture spectrum was degraded to match the {\it S/N} of the brightest spaxel of the cube. The best-fit LOSV and velocity dispersion for each spectrum are shown in the blue text boxes. We see that for an uncorrected spectrum it is not possible to obtain a reliable fit due to the wiggles, having a difference $\sim$145 km s$^{-1}$ with the LOSV of the integrated spectrum, $3\times$ larger than the estimated uncertainty. While, for both the spectrum corrected using {\sc WICKED} and the code by \citet{Perna2023} a good fit can be achieved, with both being consistent within uncertainties with the LOSV of the integrated spectrum (wiggles-free). We note that using {\sc WICKED} we obtain a best-fit value closer to the ``true'' value of $119.9 \pm 2.7$ for the integrated spectrum, and an uncertainty $\sim 45\%$ smaller than for the spectrum corrected with the code by \citet{Perna2023}, which is probably due to the superior performance of {\sc WICKED} at removing the wiggles.
\textit{Right:} We display the residuals between each spectrum and the best-fit. We can see that the residuals for the spectrum corrected using {\sc WICKED} (red) are flat and mostly dominated by outliers left during the data reduction with the JWST pipeline, while for the spectrum corrected using the code by \citet{Perna2023} there are plenty of residual wiggles. 
}
\label{fig:LOSV}
\end{figure*}

The left panels in Figure~\ref{fig:LOSV} show the best-fit model with {\sc pPXF} (blue) for the four different spectra, and the right the residuals. The degraded aperture spectrum (black) has a best-fit line-of-sight velocity of $V_{LOSV} = -119.4 \pm 2.9$ km s$^{-1}$. The un-degraded aperture spectrum (with a {\it S/N} of $150$) has an LOSV of $V_{LOSV} = -114.9 \pm 2.7$ km s$^{-1}$, perfectly matching the Gaia DR3 catalog value for this object \citep{GaiaEDR3_phot}. It is clear in Figure~\ref{fig:LOSV} that for the uncorrected data cube (gray), {\sc pPXF} cannot find a good fit due to the pronounced wiggles in the continuum, resulting in a poorly fit LOSV of $V_{LOSV} = 25.2 \pm 53.6$ km s$^{-1}$, clearly inconsistent with the star’s LOSV. For datacubes corrected for wiggles, both {\sc WICKED} (red) and the code by \citet{Perna2023} (yellow) yield reliable {\sc pPXF} fits, with LOSV values of $V_{LOSV} = -121.0 \pm 4.4$ km s$^{-1}$ for {\sc WICKED} and $V_{LOSV} = -122.5 \pm 6.4$ km s$^{-1}$ for the code by \citet{Perna2023}, both consistent with the aperture-extracted spectrum’s LOSV within error. We note that the LOSV uncertainty for the {\sc WICKED}-corrected spectrum is about 50\% smaller than that obtained with the \citet{Perna2023} method, likely due to {\sc WICKED}'s superior wiggle removal. 
The velocity dispersion for this star is below NIRSpec's instrumental resolution. However, the degraded aperture-spectrum and the datacubes corrected with {\sc WICKED} and the \citet{Perna2023} method all agree on a velocity dispersion of $0.6 \pm 0.0$ km s$^{-1}$. In contrast, the uncorrected datacube shows an unrealistic value of $449.4 \pm 101.2$ km s$^{-1}$, underscoring the impact of wiggles on determining reliable velocity dispersions. Without correction, {\sc pPXF} misinterprets broad wiggle features as actual spectral features, leading to inaccurate fits.

The right panels of Figure~\ref{fig:LOSV} show the residual between the data and their best {\sc pPXF} fit. The method of \citet{Perna2023} is less effective. The residuals for the spectrum corrected using {\sc WICKED} are mostly flat and dominated by remaining outliers from the JWST pipeline, while for the spectrum corrected using \citet{Perna2023} there are plenty of residual wiggles in the spectrum. We also note, that the spectrum corrected using the method by \citet{Perna2023} also leaves a ``concave-like'' shape when compared to the aperture-extraction spectrum. This effect is not visible in the bottom panel of Fig~\ref{fig:LOSV} because the added polynomial using during the {\sc pPXF} fit. This change in the continuum was not visible for the spectrum corrected using {\sc WICKED}.

Finally, without wiggle correction, single-pixel kinematics measurements for the spectrum similarly affected by wiggles to this M-star would be impossible, and in other cases would make it significantly challenging and bias the results. However, {\sc WICKED} can effectively remove wiggles, allowing kinematic analysis at the single-pixel level, with {\sc WICKED} providing better results than the method used by \citet{Perna2023}. 

\section{Stellar \& Gas Kinematics: Study case of NGC5128}
\label{sec:Kinematics_CenA}
In the following section, we present the stellar \& gas kinematics for the NIRSpec observations of NGC5128, hereafter Centaurus A. We showcase the kinematics of Centaurus A as a practical example of the science that can be achieved after correcting for ``wiggles'' with {\sc WICKED}. Centaurus A serves as a perfect example for this, since its nuclear kinematics has been well studied previously \citep[e.g.,][]{Silge2005,Marconi2006,Neumayer2007,Cappellari2009}, making it a perfect benchmark for comparing our results. 

We focus our kinematic analysis only in the F170LP observations, since it contains the $2.30\:\mu m$ (2-0) $^{12}$CO bandhead and two of the other prominent $^{12}$CO series, commonly used for the dynamical modeling of black hole masses, as well as hydrogen molecular and hydrogen recombination lines Pa${\alpha}$, Br${\gamma}$ and Pf${\delta}$. We removed wiggles from the F170LP observation using {\sc WICKED}, with a Fourier ratio of $3.5\sigma$ leading to 120 spaxels flagged. 

The spectrum of Centaurus~A shows prominent hydrogen recombination lines in the wavelength range of the F170LP , such as Pa$_{\alpha}$, Br$_{\gamma}$ and Br$_{\beta}$. Hydrogen recombination lines can be excited by the radiation field of the central AGN in the narrow-line region, but they can also come from radiation from ongoing star formation. In this wavelength range there are also a few important molecular hydrogen lines at $1.95\:\mu m$ 1-0 S(3) ,$2.03\:\mu m$ 1-0 S(2),$2.12\:\mu m$ 1-0 S(1). There are also several different ion lines such as [Si Vi], [OIII], [MgII], etc. In this work we do not intend to make an exhaustive study of the kinematics of the different emission lines, but extract the overall kinematic properties of the gas and stars in Centaurus~A and compared them with previous results of \citet{Neumayer2007} for the gas kinematics and \citet{Cappellari2009} for the stellar kinematics. Please refer to those works for a more in-depth review of the kinematics of Centaurus~A, as well as \citet{Neumayer2010}. 

In order to increase the {\it S/N}, we spatially bin the {\sc WICKED}-cleaned data cube using the Voronoi {\sc python} package {\sc VorBin} \citep{Cappellari2003} to achieve a {\it S/N} of 100, resulting in 515 bins, with most of the bins inside the $0.5\arcsec$ consisting of a single spaxel. We fited the spectrum of the individual bins using ({\sc pPXF}) \citep{Cappellari2017} in the spectra range of $1.7-3.16 \mu m$, with the synthetic high-resolution stellar templates from the Phonix library \citep{phoenixlib}, in the same manner as described in subsection~\S~\ref{subsec:LOSV_WICKED}. We simultaneously fit the hydrogen recombination lines, molecular lines, and stellar absorption in {\sc pPXF}, assigning separate kinematic components to each. For the gas, we fit the first two velocity moments, while for the stars, we also include the third and fourth Gauss-Hermite moments (h$_3$ and h$_4$). The typical uncertainties in LOSV and velocity dispersion for individual bins are 6 km s$^{-1}$ and 5 km s$^{-1}$, respectively. As mentioned in \citet{Cappellari2009}, the nuclear non-thermal component in Centaurus~A dominates the total flux at radii $\lesssim 0.2\arcsec$ and dilutes the stellar features in the spectrum. Therefore, it is advisable to use a fixed set of stellar templates to correctly model the spectrum and obtain reliable kinematics in the nuclear region \citep{Cappellari2009}. We define a set of stellar templates selected from the fit of an annular spectrum of Centaurus~A, to exclude the nuclear non-thermal continuum. Based on the surface brightness profile for Centaurus~A of \citet[Fig. 6]{Cappellari2009}, the non-thermal and stellar continuum  are equal at a radius of $\sim 1.0\arcsec$, so our annular spectrum is extracted between a radii of $1.0-1.3\arcsec$. The pPXF fit for the annular spectrum resulted in 9 Phoenix templates with temperature ranging from $2800-4600$ K and metallicity $-4.0$ to $-0.5$ [M/H].

\begin{figure*}[h!]

\includegraphics[width=.49\textwidth]{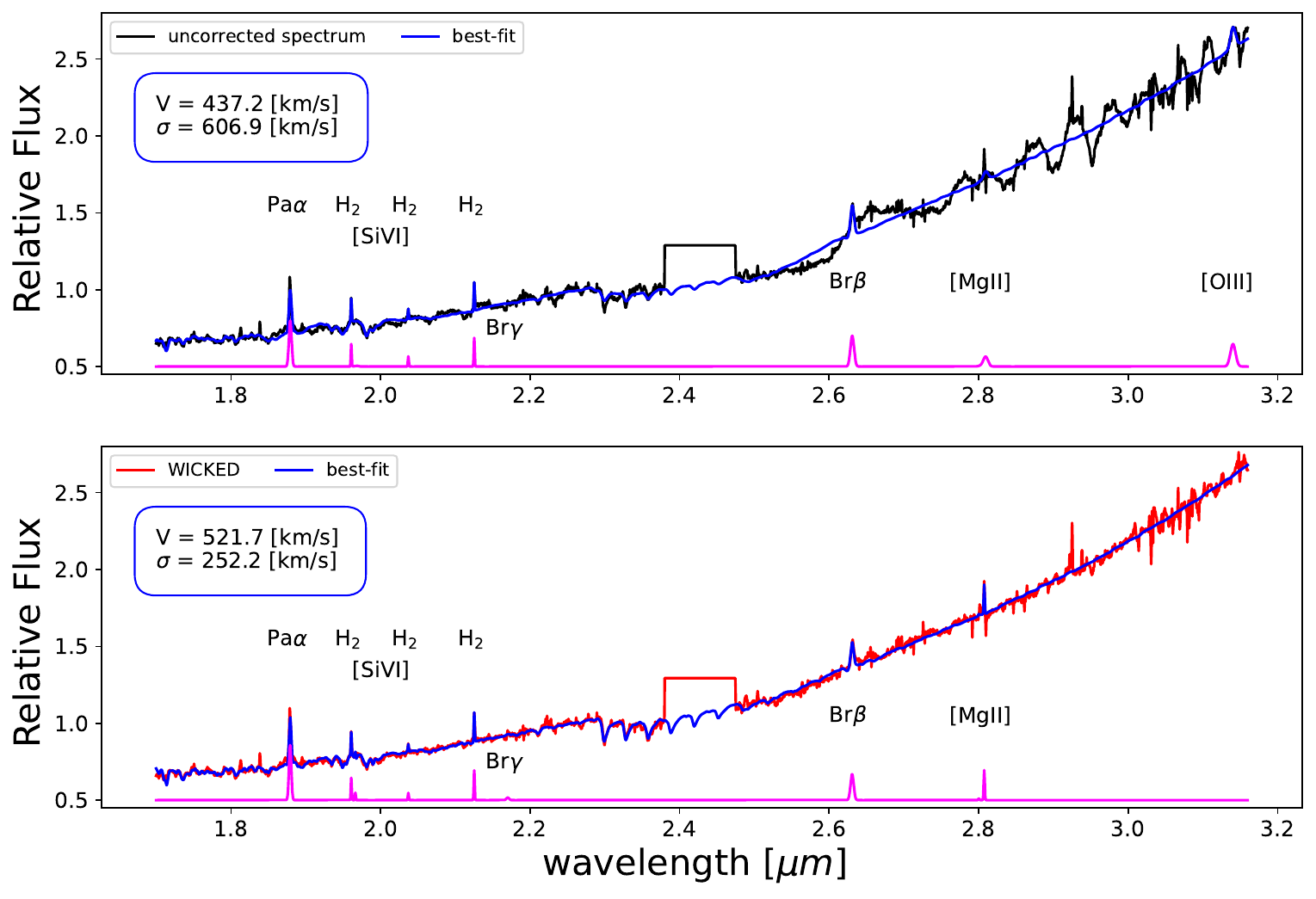}
\includegraphics[width=.5\textwidth]{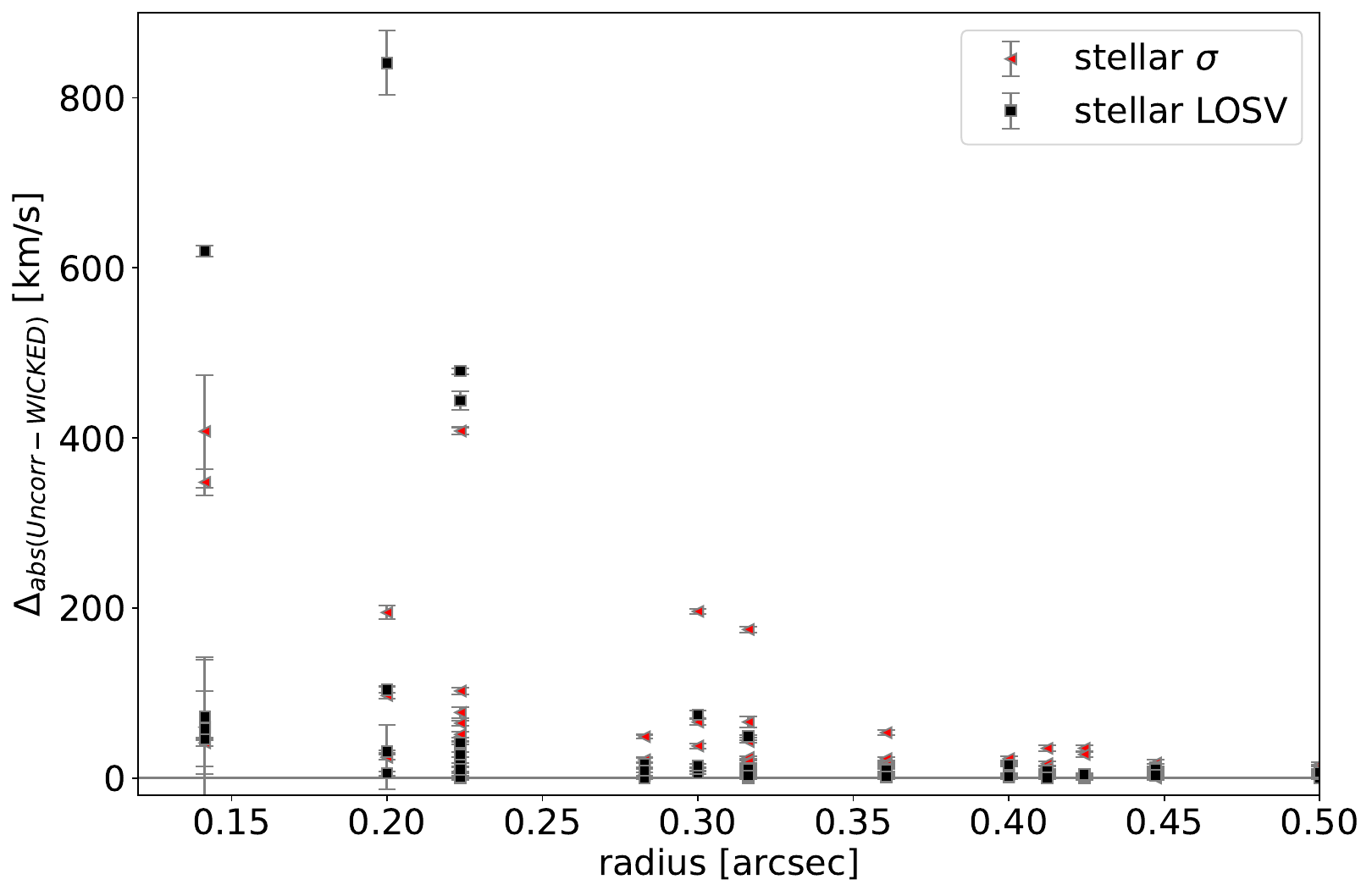}
\caption{{\it Left}: Comparison between the best {\sc pPXF} fit (blue) for an off-center spaxel of the F170LP data of Centaurus~A (at a radius of $0.15\arcsec$) of an uncorrected cube (top,gray) and one corrected using {\sc WICKED} (bottom,red). The wiggles in the uncorrected cube bias the ability of {\sc pPXF} to get a good velocity dispersion and cause the code to incorrectly identify emission lines (like the [OIII] line ). The wiggles also impacts the LOSV, having a value in disagreement with the stellar rotation of $V=531 \pm \Delta V\approx 25$ km s$^{-1}$ found in \citet{Cappellari2009}. {\it Right}: Absolute difference between the LOSV (black squares) and velocity dispersion (red triangles) for the binned spectra within $0.5\arcsec$, comparing uncorrected and {\sc WICKED}-cleaned spectra. The mean absolute difference for the LOSV (black dashed line) is $\sim$181 km s$^{-1}$, while for the velocity dispersion, it is $\sim$104 km s$^{-1}$. These values are $\sim$17$\times$ and $\sim$30$\times$ larger, respectively, than the propagated uncertainties of $\sim$8 km s$^{-1}$ } 
\label{fig:example_CenA}
\end{figure*}

An example spectrum for an off-center spaxel at a radius of $0.15\arcsec$ from the center for the uncorrected cube (top, black) and the cube corrected with {\sc WICKED} (bottom, red) are shown in Figure~\ref{fig:example_CenA}.Their best {\sc pPXF} fit is shown in blue for both spectra and their best-fit LOSV and $\sigma$ in the blue box. In this example, we can see the effect that the wiggles have in determining the LOSV and $\sigma$ and also in determining emission lines with {\sc pPXF}. The wiggles in the uncorrected spectrum bias the best-fit LOSV to a value much larger than the $531 \pm \Delta V = 20$ km s$^{-1}$ reported in \citet{Cappellari2009} inside the $1.5\arcsec$. The LOSV value for the spectrum corrected with {\sc WICKED} is in great agreement with this value. Similarly for the velocity dispersion $\sigma$, since the model tries to fit the wide features of the wiggles as if they were stellar features, the fitted $\sigma$ is biased high for the uncorrected cube, to values unrealistically higher than the maximum value of $\sim 165$ km s$^{-1}$ from \citet{Cappellari2009}. The wiggles also affect the ability to detect emission lines in the spectrum. For the uncorrected spectrum {\sc pPXF} confuses a wiggle at $3.13 \mu m$ as the [OIII] line, while in the corrected spectrum with {\sc WICKED} it is not present. In general, for spaxels affected by wiggles (radii $\leq0.4\arcsec$), we observe an average difference of $\sim$181 km s$^{-1}$ and $\sim$104 km s$^{-1}$ for the LOSV and velocity dispersion between the uncorrected cube and the {\sc WICKED}-cleaned cube, respectively (right-panel in Figure~\ref{fig:example_CenA}). These difference is about 36$\times$ and 17$\times$ the average uncertainty for the LOSV and velocity dispersion, respectively. 

\begin{figure*}

    \includegraphics[width=1\textwidth]{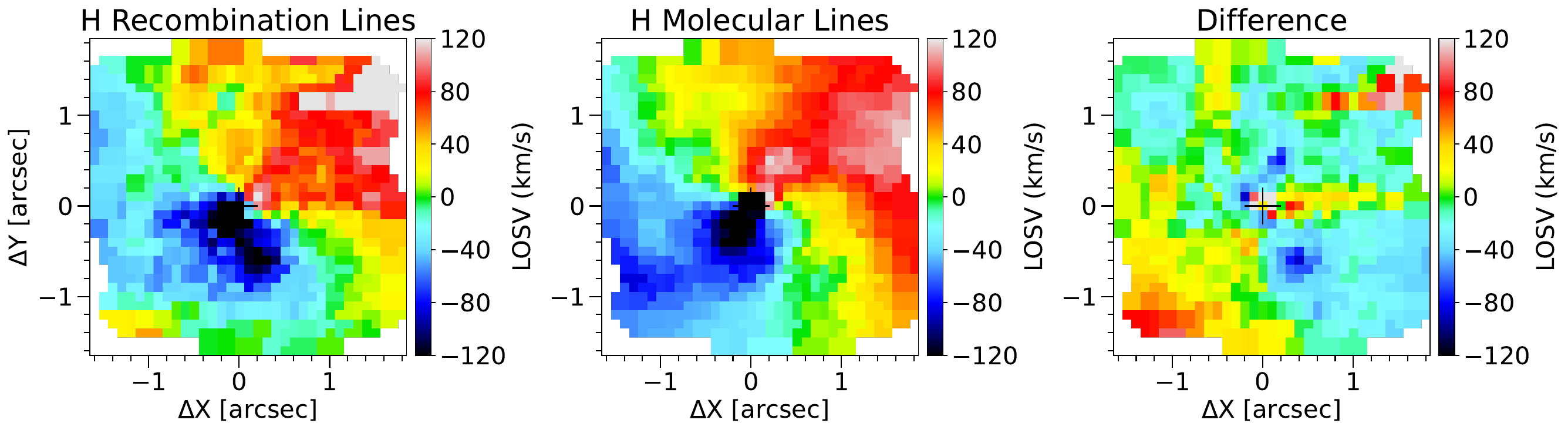}
\caption{Voronoi binned LOSV velocity maps for the Hydrogen recombination (\it{left}) \& molecular lines (\it{middle}) for the F170LP NIRSpec observations of Centaurus~A. The difference velocity map between the Hydrogen recombination \& molecular lines (right panel) shows a velocity gradient along the radio-jet consistent with the finding of \citet{Neumayer2007}. The maps are oriented such that North is up and east is to the left.}
\label{fig:gas_kinematics}
\end{figure*}

The extracted gas kinematics for the F170LP cube corrected with {\sc WICKED} for the hydrogen recombination lines and hydrogen molecular lines and the difference between the two are shown in Figure~\ref{fig:gas_kinematics}. The velocity map for the molecular hydrogen (middle panel in Fig~\ref{fig:gas_kinematics}) is smooth and symmetric with a clear regular rotational pattern and a maximum rotation velocity of $\Delta V \approx 123$ km s$^{-1}$. The velocity map also shows the same twist in the major axis of rotation reported in \citet{Neumayer2007}. Overall the kinematics of the molecular hydrogen is in good agreement with the results from \citet{Neumayer2007}. 
The kinematics for the hydrogen recombination lines on the other hand also shows a clear rotation, but it seems to be somehow influenced by the Centaurus~A's radio-jet. The difference in the velocity pattern between the two is clear when we plot their difference (right panel, Fig~\ref{fig:gas_kinematics}. Their difference shows a dip in velocity in the region $(0.5\arcsec,-0.5\arcsec)$ which is close to the knot in the radio jet shown in \citet{Neumayer2007}. The velocity map for Br$\gamma$ from \citet{Neumayer2007} aligns well with our hydrogen recombination line results, which include Br$\gamma$.

\begin{figure*}
    \includegraphics[width=1\textwidth]{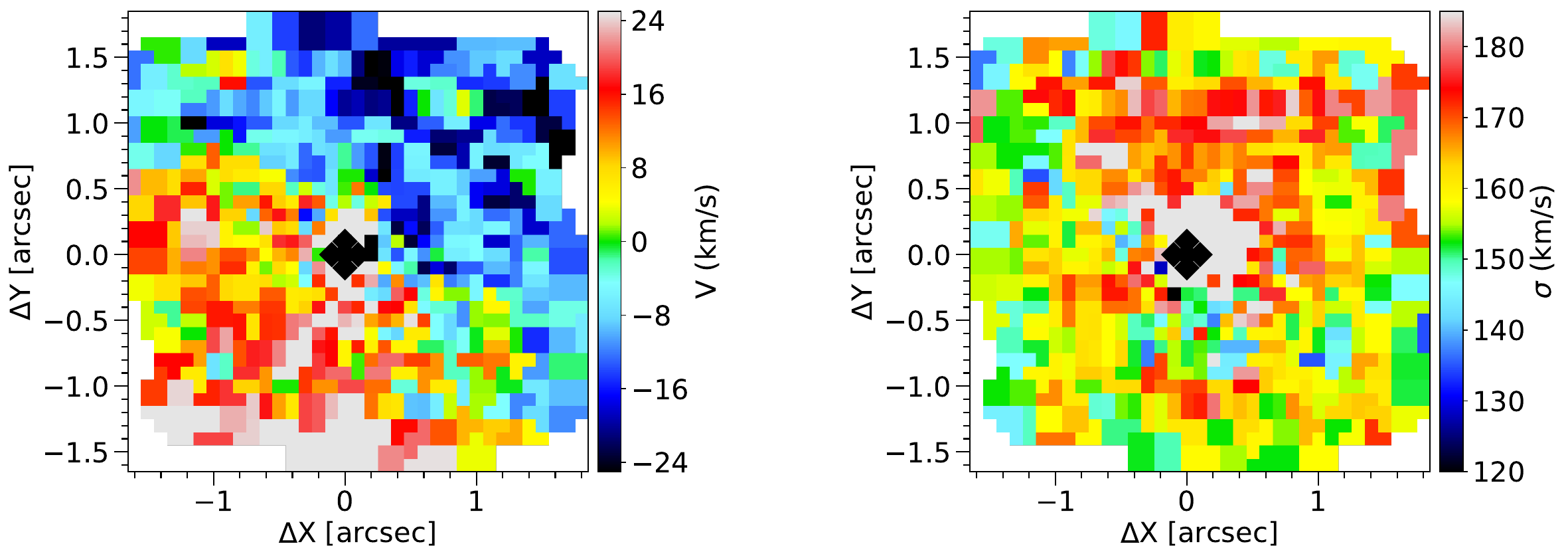}
    \includegraphics[width=1\textwidth]{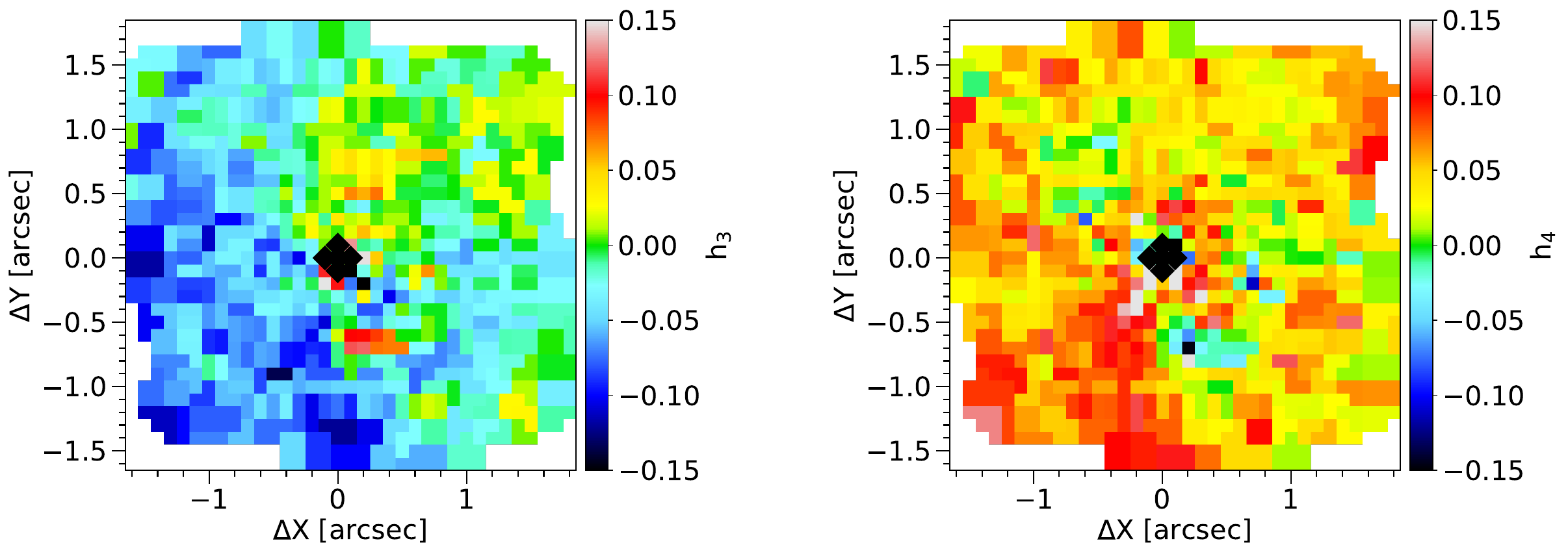}
\caption{Voronoi binned stellar velocity maps for the F170LP NIRSpec observations. The four panels show the line-of-sight velocity (V), the velocity dispersion $\sigma$, and the the first two Gaussian-Hermite moments, h$_3$ and h$_4$. The black diamonds show the 5 masked spaxels where a stellar fit is not possible due to the strong non-thermal AGN continuum. The orientation is the same that in Figure~\ref{fig:gas_kinematics}.}
\label{fig:stellar_kinematics}
\end{figure*}

Figure~\ref{fig:stellar_kinematics} shows the stellar LOSV, velocity dispersion and the first two Gaussian-Hermite moments. The strong non-thermal continuum in the nuclear region of Centaurs~A almost completely dilutes every stellar feature in the spectrum inside a radius of $\sim 0.1\arcsec$, therefore we exclude those bins (five bins) and are shown as black diamonds in Figure~\ref{fig:stellar_kinematics}. The nuclear stellar rotation exhibits a counter-rotation of approximately $180^\circ$ relative to the molecular hydrogen, and with a much slower rotation. This stellar counter-rotation has already been shown in \citet{Cappellari2009} and suggests that the in-falling gas has not been able to produce a large fraction of stars yet. While the stellar rotation and the profile of the velocity dispersion matches the pattern found in \citet{Cappellari2009}, our velocity dispersion is in average $\sim 10\%$ and $5\%$ higher at a radius of $0.2\arcsec$ and $0.4\arcsec$ respectively. This could be due to the superior sensitivity of NIRSpec compared to the SINFONI data used \citet{Cappellari2009} and the higher {\it S/N}, with a minimum {\it S/N} of 100 and a {\it S/N} of $\geq 400$ for the bins inside the $0.25\arcsec$.

The h$_3$ field shows an anti-correlation with respect to the LOSV as observed in other early-type galaxies, and it is associated with a disc structure \citep[e.g][]{Krajnovic2008}. The expected anti-correlation between the LOSV and the h$_3$ is more clear in the F170LP than in the SINFONI 100 mas data from \citet{Cappellari2009}, and it is in the expected range of values for the $h_3 - V/\sigma$ relation for slow-rotator galaxies such as Centaurus~A in the SAURON sample from \citet{Krajnovic2008}. \citet{Cappellari2009} also reports a central symmetric structure in the inner $\sim 0.5\arcsec$ which suggests a possible template mismatch. The template mismatch would also affect the value of $h_4$ of \citet{Cappellari2009}, which are more susceptible to template mismatch \citep{Krajnovic2008}. Our $h_4$ fields show a flat structure, with a small radial symmetry and a mean value of $<h_4> \approx 0.04$. The higher overall $h_4$ may also be linked to the increased velocity dispersion we find compared to \citet{Cappellari2009}.

\section{Conclusion}
\label{sec:Conclusion}
In this work we presented a {\sc Python} tool called {\sc WICKED} to remove the point-spread function artifacts know as ``wiggles'' from the different NIRSpec gratings in the IFU mode. We showed that spaxels in the data cube affected by these artifacts can be distinguished from the rest by analyzing their Fast Fourier Transform \S~\ref{subsec:flagging_pixels}. {\sc WICKED} removes the wiggles in the spectrum by taking advantage of the correlation between their frequency $f_\lambda$ and wavelength presented in \citet{Perna2023}, but with severe modification leading to a better removal of wiggles and preservation of the continuum shape and emission/absorption lines in the spectrum.  We performed different tests to the wiggle-free stellar spectrum of an A-star and an M-star to quantify the effectiveness of {\sc WICKED} at identifying spaxels with wiggles versus Gaussian random noise, how well it preserves the equivalent width of absorption lines, the shape of the spectrum, the line-of-sight velocity and velocity dispersion. The results of these tests are the following:

\begin{itemize}
    \item The frequency of the wiggles in the spectrum depends on the brightness of the source and the dither pattern used during the observations. However, for a specific object they have a defined range of frequencies that can be used to identify wiggles from the rest of features in the spectrum using the Fast Fourier Transform. For the spectrum of the A-star used in our test, the mean amplitude of the Fourier spectrum at these frequencies is grater than the mean amplitude for the rest of the Fourier spectrum by a factor of $\geq 3\sigma$ up to a {\it S/N} of $\sim15$, and $\geq 1\sigma$ at {\it S/N} of $\sim 8$. 
    
    \item We model the spectrum of flagged spaxels using a combination of an aperture spectrum, an annular spectrum, a power law, and a second-degree polynomial. These components are optimized by minimizing the chi-square of the fit at each spaxel. The two spectral templates help account for spatial variations in the cube and model different spectral features. Since many physical processes, especially at shorter wavelengths, can be represented by a power law, we include it to capture changes in the spectral shape that can not be modeled using just the templates. The second-degree polynomial helps handle any remaining mismatches, such as common rises or drops at the spectrum edges or large bumps that sometimes show up (probably also PSF artifacts). We chose this approach instead of using a high-degree polynomial, as done in \citet{Perna2023}, because high-degree polynomials can behave unpredictably and may mimic sinusoidal patterns, partially fitting the wiggle amplitudes and impacting the ability to remove wiggles. 
    
    \item The equivalent width of the spectrum corrected with {\sc WICKED} for multiple lines shows a maximum relative difference of $5\%$ from the true value across different ${\it S/N}$ levels. On average, this is less than half the difference seen with the method by \citet{Perna2023} and  $4\times$ smaller than the differences in the uncorrected spectrum. Differences in the overall spectral shape are up to $3.5\times$ smaller compared to the uncorrected spectrum at {\it S/N} $\geq 200$ and about $2.5\times$ smaller at {\it S/N} $\leq 100$. Compared to the spectrum corrected with \citet{Perna2023}, the difference is about $1.5\times$ smaller across all {\it S/N} levels. While the correction from \citet{Perna2023} leaves large regions of residual wiggles, especially at the spectrum edges, the spectrum corrected with {\sc WICKED} only shows narrow “spikes” residuals around some absorption lines, consistent with the $5\%$ difference mentioned earlier.
    
    \item Wiggles significantly impact the ability to perform single-pixel level kinematics, since they get mistaken as spectral features, producing artificially large velocity dispersions, and they can also get mistaken as broad emission lines. {\sc WICKED} showed a better performance at recovering the true LOSV of a the M-star used in our test, as well as reducing its uncertainty by a factor of $\sim 50 \%$ compared to the method by \citet{Perna2023}. 

\end{itemize}

We also applied {\sc WICKED} to correct for wiggles and analyze the nuclear gas and stellar kinematics of NGC~5128 or Centarurus~A. This serve as a real example of the type of science that can be unlocked when removing these artifacts. We showed how the broad wiggles in the uncorrected F170LP cube of Centaurus~A, lead {\sc pPXF} to wrongly identified emission lines in the spectrum and impacted the kinematics, since {\sc pPXF} would fit the wiggles as broad spectral features. The resulting gas and stellar kinematics of Centaurus~A are in good agreement with previous works \citep{Neumayer2007,Cappellari2009}, showing a regular rotation for the molecular hydrogen and for the hydrogen recombination lines but with some distortion southwest of the center close to a knot in the radio jet \citep{Neumayer2007}. The stellar component shows a slow disk type of rotation that is counter-rotating by $\sim 180^\circ$ respect to the molecular hydrogen. The stars also show a increase in velocity dispersion close to the center up to a radius of $\sim 0.1\arcsec$ where the strong non-thermal component in the spectrum completely vanishes the CO bandhead at $2.3\mu m$ impacting our ability to obtain a reliable fit. The first Gaussian-Hermite moment h$_3$ shows a clear anti-correlation with respect to the stellar LOSV as expected for early-type galaxies \citep{Krajnovic2008}, with values similar to the SINFONI 100mas data in \citet{Cappellari2009} but with a more clear anti-correlation. The second Gaussian-Hermite moment h$_4$ shows a mean value of h$_4 = 0.04$ and is quite symmetrical with respect to the center. The value of h$_4$ is higher than reported in \citet{Cappellari2009} but within expectation for slow-rotator galaxies \citep{Krajnovic2008}. The higher h$_4$ could also explain why we get in average a $\leq$10\% higher velocity dispersion  inside the $0.2\arcsec$ than \citet{Cappellari2009}. We believe that the disagreement is due to the template-mismatch mentioned in \citet{Cappellari2009} which can impact the determination of h$_4$ \citep{Krajnovic2008}, and the superior sensibility of JWST NIRSpec compared to their SINFONI data, which is why we also observe a more clear anti-correlation between the stellar LOSV and the h$_3$ in our data. 

The good agreement between the LOSV from the {\sc WICKED}-cleaned spectrum and the aperture spectrum of the M-giant star J15395077 (Section~\ref{subsec:LOSV_WICKED}), along with the overall consistency of NGC~5128’s gas and stellar kinematics with previous studies, suggests that the kinematics recovered from the {\sc WICKED}-cleaned data cube closely represent the true values. Based on this, we find that the uncorrected spectra show significant differences in both LOSV and velocity dispersion compared to the {\sc WICKED}-cleaned spectra, with discrepancies on average of $\sim$30$\times$ and $\sim$12$\times$ the typical propagated uncertainties, with the largest outlier reaching $\sim$138$\times$ and $\sim$93$\times$, respectively (see Figure\ref{fig:example_CenA}). These highlight the importance of correcting for wiggles, and that failing to do so can introduce substantial biases in kinematic measurements, far exceeding expected uncertainties.

Wiggles impact significantly the possibility of doing single-pixel science across the different gratings of JWST NIRSpec in the IFU mode. The user-friendly aspects of {\sc WICKED} combined with its capability for correcting these artifacts across different signal-to-noise ratios presents a practical solution for the community to exploit the NIRSpec data, avoiding spatial binning to remove wiggles. 
{\sc WICKED} has already been applied for some datasets that will be presented in the future works. We applied {\sc WICKED} to clean the data for the ``Revealing Low-Luminosity Active Galactic Nuclei'' (ReveaLLAGN) sample, where we will present the stellar \& gas kinematics in a future paper ({\it Dumont et al 2025 (in prep.)}). It has also been used for cleaning the data for constructing the PSF model in {\it Ohlson et al 2025 (in prep.)} and for cleaning the spectra of a couple of high-redshift Quasars in {\it Wolf et al 2025 (in prep.)}. 

The {\sc Python} package for {\sc WICKED} is freely available and can be downloaded from the {\sc GitHub} repository \hyperlink{}{https://github.com/antoinedumontneira/WiCKED}

\section*{Acknowledgement}

LCH was supported by the National Science Foundation of China (11991052, 12233001), the National Key R\&D Program of China (2022YFF0503401), and the China Manned Space Project (CMS-CSST-2021-A04, CMS-CSST-2021-A06. 

\bibliography{References}
\bibliographystyle{aasjournal}

\end{document}